\documentstyle{mn2e}
\input psfig

  \makeatletter
  \ifx\CUP@mtlplain@loaded\undefined  
  \makeatother  
    
  \else
    
  \fi

\loadboldmathitalic 
\loadboldgreek      
  
\makeatletter
\ifx\CUP@mtlplain@loaded\undefined  
  \newfont\bit{cmbxti10 at 9pt}
\else
  \newfont\bit{mtbxti10 at 9pt}
\fi
\makeatother

\def\LaTeX{L\kern-.36em\raise.3ex\hbox{a}\kern-.15em
    T\kern-.1667em\lower.7ex\hbox{E}\kern-.125emX}

\newcommand{\gsim}{\mathrel{\hbox{\rlap{\lower.55ex \hbox {$\sim$}}
                   \kern-.3em \raise.4ex \hbox{$>$}}}}
\newcommand{\lsim}{\mathrel{\hbox{\rlap{\lower.55ex \hbox {$\sim$}}
                   \kern-.3em \raise.4ex \hbox{$<$}}}}

\newcommand{\fe}{${\rm f}^{\rm e}$\ }
\newcommand{\pe}{${\rm p}^{\rm e}_{1}$\ }
\newcommand{\fo}{${\rm f}^{\rm o}$\ }
\newcommand{\foro}{${\rm f}^{\rm o}$/${\rm r}^{\rm o}_{1}$\ }
\newcommand{\ro}{${\rm r}^{\rm o}_{1}$\ }
\newcommand{\re}{${\rm r}^{\rm e}_{1}$\ }
\newcommand{\fr}{$|f_{\rm r}|$\ }

\title[Waves in accretion discs]{The excitation, propagation and 
  dissipation of waves in accretion discs: the non-linear axisymmetric case}

\author[M. R. Bate et~al.]
  {M. R. Bate,$^{1,2}$\thanks{E-mail: mbate@astro.ex.ac.uk}
  G. I. Ogilvie,$^1$
  S. H. Lubow,$^{1,3}$
  and J. E. Pringle$^{1,3}$\\
  $^1$Institute of Astronomy, University of Cambridge, Madingley Road,
    Cambridge CB3 0HA \\
  $^2$School of Physics, University of Exeter, Stocker Road,
    Exeter EX4 4QL \\
  $^3$Space Telescope Science Institute, 3700 San Martin Drive,
    Baltimore, MD 21218, USA
}

\date{\today}

\begin{document}

\maketitle

\begin{abstract}
  We analyse the non-linear propagation and dissipation of
  axisymmetric waves in accretion discs using the ZEUS-2D
  hydrodynamics code.  The waves are numerically resolved in the
  vertical and radial directions.  Both vertically isothermal and
   thermally stratified accretion discs are considered.  The waves are
  generated by means of resonant forcing and several forms of forcing
  are considered.  Compressional motions are taken to be locally
  adiabatic ($\gamma = 5/3$).  Prior to non-linear dissipation, the
  numerical results are in excellent agreement with the linear theory
  of wave channelling in predicting the types of modes that are
  excited, the energy flux by carried by each mode, and the vertical
  wave energy distribution as a function of radius.  In all cases,
  waves are excited that propagate on both sides of the resonance
  (inwards and outwards).  For vertically isothermal discs, non-linear
  dissipation occurs primarily through shocks that result from the
  classical steepening of acoustic waves.  For discs that are
  substantially thermally stratified, wave channelling is the primary
  mechanism for shock generation.  Wave channelling boosts the Mach
  number of the wave by vertically confining the wave to a small cool
  region at the base of the disc atmosphere. In general, outwardly
  propagating waves with Mach numbers near resonance ${\cal M}_{\rm r}
  \ga0.01$ undergo shocks within a distance of order the resonance
  radius.
\end{abstract}


\section{Introduction}

Gaseous discs play a significant role in the evolution of binary star
and planetary systems.  Generally, the tidal forces from stellar or
planetary companions distort the discs from an axisymmetric form.
However, where resonances occur in the discs, the tidal forces also
generate waves that transport energy and angular momentum.  The
resulting resonant torques cause orbital evolution of the perturbing
objects (e.g. Goldreich \& Tremaine 1980; Lin \& Papaloizou 1993;
Lubow \& Artymowicz 1996).  The waves also cause the discs to evolve
since they transfer their energy and angular momentum to the disc in
the locations where they damp.  Damping may be provided by shocks,
radiative damping (Cassen \& Woolum 1996), or other viscous damping
mechanisms.

Waves in discs have also been considered as possible explanations of
quasi-periodic variability in the X-ray emission of systems involving
accretion discs around black holes (e.g. Kato \& Fukue 1980; Nowak \&
Wagoner 1991; Kato 2001).

Initial studies of wave propagation approximated the disc as
two-dimensional and ignored the effects of vertical structure
(perpendicular to the disc plane).  Goldreich \& Tremaine (1979)
developed a two-dimensional linear theory for resonant tidal torques
and the associated wave propagation.  Later studies of the non-linear
case found that the torques were within a few percent of those
predicted by the two-dimensional linear formula (Shu, Yuan \& Lissauer
1985; Yuan \& Cassen 1994). For a thin disc that is vertically
isothermal and has an isothermal thermodynamic response ($\gamma =
1$), the only wave excited in the disc has a two-dimensional
structure. The wavefronts are perpendicular to the disc plane and the
motion is purely horizontal (parallel to the disc plane).  Thus, a
two-dimensional treatment is valid in a thin disc if both the vertical
structure of the disc and its thermodynamic response are locally
isothermal.  However, this is not realistic for most discs, such as
those in cataclysmic variables, circumstellar and circumbinary discs
around pre-main-sequence stars, and protoplanetary discs.  In such
discs, the waves are no longer two-dimensional and the vertical
structure of the disc must be taken into account.

A study of three-dimensional wave propagation was made using
linearised numerical simulations (Lin, Papaloizou \& Savonije
1990a,b).  Unfortunately, the limited range of physical parameter
space that was covered led to the interpretation that waves in a
vertically thermally stratified disc are refracted upwards into the
atmosphere where they dissipate via shocks.  Later, it was realised
that the linear problem could be solved semi-analytically (Lubow \&
Pringle 1993; Korycansky \& Pringle 1995; Lubow \& Ogilvie 1998;
Ogilvie \& Lubow 1999).  These studies found that the propagation of
waves in a thin disc is similar to the propagation of electromagnetic
waves along a waveguide (Jackson 1962; Landau \& Lifshitz 1960).  The
motions in the disc can be described in terms of modes.  For each
mode, the vertical wave structure is determined in detail at each disc
radius. The horizontal variations of the mode are treated by means of
a WKB approximation in which the horizontal (radial) wavenumber is
determined as a function of wave frequency. A flux conservation law
determines the mode amplitude as a function of radius.  Depending on
the vertical structure of the disc and the thermodynamic response of
the gas, the wave energy may be channelled towards the surface of the
disc, towards the mid-plane of the disc, or occupy the entire vertical
extent as it propagates in radius.

Ideally, one would like to determine the non-linear, non-axisymmetric
response of a variety of disc models (vertically isothermal and
thermally stratified) to resonant forcing. Unfortunately, this is a
formidable numerical problem at present. Instead we begin, in this
paper, with an analysis of the non-linear axisymmetric response.  The
axisymmetric case offers the major advantage of being expressed as a
two-dimensional problem, which is currently accessible numerically.
Although axisymmetric waves only transport energy and not angular
momentum, they should resemble the response of the disc to
low-azimuthal number tidal forcing, as arises in binary star systems.
The local physics of the waveguide
is determined within a region whose size is a few times the 
semi-thickness of the disc.  On this scale a non-axisymmetric 
wave of low azimuthal wavenumber is almost indistinguishable 
from an axisymmetric wave.  This explains why the local physics 
of the waveguide can be studied within the shearing-sheet 
approximation (Lubow \& Pringle 1993) and why the azimuthal
wavenumber appears in the dispersion relation only through a 
Doppler shift of the wave frequency.

Consequently, in this paper, we study the non-linear propagation and
dissipation of axisymmetric, but fully resolved, waves in accretion
discs.  The purpose of the paper is threefold.  First, we wish to
determine, using non-linear hydrodynamic calculations, whether or not
the semi-analytical linear theory provides an accurate description of
wave propagation in accretion discs.  Secondly, although linear theory
can predict where non-linearity is likely to become important, shocks
and the process of energy deposition (and angular momentum deposition
with non-axisymmetric waves) cannot be handled.  We aim to determine
how and where dissipation occurs through shocks and how accurately
this can be predicted from the linear analysis.  Finally, we wish to
determine how well non-linear hydrodynamic calculations can model the
problem and what resolution is required.

The outline of this paper is as follows.  In Section 2, we briefly
review the ZEUS-2D hydrodynamic code that is used to obtain the
non-linear results.  In Section 3, we discuss the structure of the
equilibrium discs that we model and the grid layout for the ZEUS-2D
calculations.  Section 4 briefly reviews the semi-analytical linear
method for solving the three-dimensional wave propagation problem,
describes our method of wave excitation, and discusses the
requirements for convergence of the numerical calculations.  Sections
5 and 6 contain the results for a vertically isothermal disc, and a
polytropic disc with a vertically isothermal atmosphere, respectively.
Intermediate discs with optical depths not much larger than unity are
considered in Section 7.  A summary of the results is contained in
Section 8.

\section{Numerical method}

The non-linear hydrodynamic calculations presented here were performed
with the ZEUS-2D code developed by Stone \& Norman (1992).  No
modifications to the code were required, except the addition of
damping boundary conditions for test purposes only (see below).

The ZEUS-2D code comes with various options for the interpolation
scheme and artificial viscosity.  We use the van Leer (second-order)
interpolation scheme.  Linear and quadratic artificial viscosities are
provided in the code, with two forms of the quadratic viscosity: the
standard von Neumann \& Richtmyer (1950) form, and a tensor form.  We
use only a quadratic artificial viscosity to handle shocks.
Calculations were performed both with the von Neumann \& Richtmyer
form and with the tensor form. We find no significant difference in
the results.  All the results presented here use the standard von
Neumann \& Richtmyer form.  The strength of the quadratic viscosity is
controlled by a parameter $q_{\rm visc}$.  Generally, we use $q_{\rm
  visc}=4$, but we present some results from calculations that were
performed with $q_{\rm visc}=0.5$.  No other type of viscosity (e.g.
one that includes a shear stress) is used.

All of the calculations presented here use reflecting boundaries.  In
order to verify that our results were not affected by reflections from
the boundaries, some calculations were performed with damping
boundaries at the inner and outer radii of the numerical grid.
Damping is provided by adding an acceleration
\begin{equation}
  \frac{\partial{\bmath v}}{\partial t} = - 2 \omega {\bmath v}
\end{equation}
to the inner or outer $10-20$ radial zones of the grid, where ${\bmath
  v}$ is the velocity of the gas and $\omega$ is the frequency of the
forcing that is used to generate the waves (see Section
\ref{excitation}).  There is no significant difference between the
results obtained using reflecting and damping boundaries.

\section{Modelling the accretion discs}

We consider the propagation and dissipation of axisymmetric ($m=0$)
waves in two types of accretion disc: a locally vertically isothermal
disc, and a polytropic disc (polytropic index $n=1.5$) that joins
smoothly on to an isothermal atmosphere.  We ignore all effects of
disc turbulence. A full description of the equilibrium structures is
given in Appendix A; we only briefly list their properties here. The
disc is described in cylindrical coordinates $(R, \phi, z)$.  As
described in Appendix A, we shall work in dimensional units, such that
the disc angular velocity is unity when the dimensionless disc radius
$R = 1$.  The discs are Keplerian, with angular velocity 
$\Omega=R^{-3/2}$, apart from small corrections.

In both discs, the value of the polytropic exponent that describes the
adiabatic pressure-density relation for the waves is $\gamma=5/3$.
For convenience in modelling, we choose discs with finite inner and
outer radii, $R_1$ and $R_2$. We also choose a scale-height that
increases linearly with radius, which requires that the sound speed
varies as $c \propto R^{-1/2}$.  The scale-height of the
vertically isothermal disc is $H \propto R$.  The polytropic disc has
a well-defined polytropic layer with a semi-thickness $H_{\rm p}$,
that smoothly changes into an isothermal atmosphere with its own
scale-height $H_{\rm i}$, above $z \approx H_{\rm p}$.  Both $H_{\rm
  p}$ and $H_{\rm i}$ are proportional to $R$.  From this point on, we
will refer to $H_{\rm p}$ simply as $H$ for the polytropic disc.  We
use discs with $H/R=0.05, 0.1$, and $0.2$.

The polytropic disc is a fair representation of an optically thick
disc in which the temperature is greatest at the mid-plane and
declines with increasing height until the photosphere is reached.
Above this height, the atmosphere is nearly isothermal.  We quote an
approximate effective optical depth at the mid-plane, $\tau$, through
the relation (e.g. Bell et~al. 1997)
\begin{equation}
  \tau={{8}\over{3}}\left({{T_{\rm m}}\over{T_{\rm i}}}\right)^4,
  \label{tau_def}
\end{equation}
where $T_{\rm m}$ and $T_{\rm i}$ are the temperatures at the
mid-plane and in the isothermal atmosphere, respectively.  For our
standard parameters (Appendix A and Section \ref{secpoly}), 
the effective optical depth is $\tau\approx25000$.

The mid-plane density profiles are power laws, $\rho(R, \phi, z=0)
\propto R^{-3/2}$, throughout the main body of the disc.  The density
drops smoothly at the edges of the discs over a distance equal to the
local value of $H$.  For the vertically isothermal discs, we adopt
$R_1=0.5$ and $R_2=10.0$.  The polytropic discs have outer radii of
$R_2=3.0$.  The value of the inner radius is $R_1=0.2$ for all but the
highest-resolution calculations, for which $R_1=0.6$.

To model axisymmetric waves in these discs in ZEUS-2D, we use
spherical polar coordinates $(r, \theta)$.  For the vertically
isothermal discs, we adopt inner and outer radial grid boundaries
$r_{\rm in}=0.35$, $r_{\rm out}=13$, and the grid encompasses 8
vertical scale-heights above and below the mid-plane 
(i.e. $\theta_{\rm min}=\pi/2-8 H/r$ and
$\theta_{\rm max}=\pi/2+8 H/r$).

For the polytropic discs, the outer grid radius is chosen
as $r_{\rm out}=4$, and 3 vertical scale-heights are modelled above
and below the mid-plane (except in Section \ref{sectau}, where
6 vertical scale-heights are modelled).  The inner grid radius $r_{\rm in}$
depends on the resolution.  We adopt $r_{\rm in}=0.15$ for all but the
highest resolution, for which $r_{\rm in}=0.6$.  

For all cases, we adopt a logarithmic radial grid in which
\begin{equation}
  (\Delta r)_{\rm i+1}/(\Delta r)_{\rm i} =
  \sqrt[N_r]{{r_{\rm out}}/{r_{\rm in}}},
\end{equation}
where $N_r$ is the number of radial grid zones.  The $\theta$ grid is
uniform with
\begin{equation}
  N_{\theta} = \frac{0.1}{(H/r)}~\frac{
  \theta_{\rm max}-\theta_{\rm min}}
  {(\Delta r)_{\rm i+1}/(\Delta r)_{\rm i} - 1}
\end{equation}
zones.  This choice means that discs with $H/r=0.1$ have grid zones
that have nearly equal radial and vertical lengths.  Discs with other
values of $H/r$ have the same number of zones per vertical
scale-height and the same radial spacing, but each zone is either
stretched or compressed vertically.

In practice, calculations of vertically isothermal discs were
performed with $N_r\times N_{\theta}$ zones of: $250 \times 111$,
$500\times 221$, or $1000\times 441$.  Calculations of the standard
polytropic
discs were computed with: $250\times 45$, $500\times 91$, $1000\times
183$, or $1156\times 365$ zones (the last of which has a resolution
equivalent to $2000\times 365$ zones but models a smaller range of
radii than the lower-resolution calculations).  The highest resolution
calculations took up to 6000 CPU hours on 440 MHz Sun Workstations
with each factor of 2 increase in resolution taking a factor of
$\approx 8$ longer (a factor of 4 from the increase in the number of 
zones and a factor of 2 from the decrease in the time step).  In some 
cases even higher resolution would be desirable, but doubling the 
resolution again is impractical.

The models are run until any transient structure disappears.
Typically, for discs with $H/r=0.1$, this takes $\approx 30$ orbital
periods at the resonant radius (usually $r = 1$, see below).  The
number of periods required is inversely proportional to $H/r$.  We
calculate the wave energy, mean Mach number, radial energy flux, and
dissipation rate by averaging these quantities over $N \approx 10$
wave periods, $P$, after the disc has completed $40$ rotations at the
resonant radius.  Let $(v_r,v_\theta)$ denote the velocity deviations
from Keplerian.  The mean (time-averaged) local wave kinetic energy density is
calculated as
\begin{equation}
\label{energyeq}
  E(r, \theta) = \frac{1}{NP}\int_{0}^{NP} \frac{1}{2}
  \rho (v_r^2 + v_\theta^2) \ {\rm d}t.
\end{equation}
The mean local Mach number is calculated as
\begin{equation}
\label{macheq}
  {\cal M}(r, \theta) = \frac{1}{NP}\int_{0}^{NP}
  \sqrt{\rho (v_r^2 + v_\theta^2) /(\gamma p)} \ {\rm d}t.
\end{equation}
The mean vertically integrated radial energy 
flux is calculated as
\begin{equation}
\label{fluxeq}
  f_{\rm r}(r) = 2 \pi \int_{\theta_{\rm min}}^{\theta_{\rm max}}
  \frac{1}{NP}\int_{0}^{NP} v_r (p-p_{\rm 0}) \ {\rm d}t
  \ r^2\sin{\theta} \ {\rm d}\theta 
\end{equation}
where $p_{\rm 0}$ is the mean pressure, which is calculated over the
previous wave period.  The dissipation rate is calculated as the
average over several wave periods of the rate, per unit volume, at
which thermal energy is deposited by the artificial viscosity.

\section{Axisymmetric wave excitation and propagation}

\label{excitation}

The radial propagation of linear axisymmetric waves in a vertically
isothermal accretion disc was analysed by Lubow \& Pringle (1993).
Subsequently, this work was extended to study the radial propagation
of waves in vertically thermally stratified discs (Korycansky \&
Pringle 1995; Lubow \& Ogilvie 1998; Ogilvie \& Lubow 1999) and
magnetized discs (Ogilvie 1998).

\subsection{The two-dimensional wave}

\label{twodimensional}

The simplest wave structure is that of a two-dimensional wave that
propagates radially through a vertically isothermal disc and has no
vertical motion (e.g. Goldreich \& Tremaine 1979).  The dispersion
relation for the axisymmetric two-dimensional wave is
\begin{equation}
\label{dispersion}
  \omega^2 = \kappa^2 + \frac{\gamma p}{\rho} k^2
\end{equation}
where $\omega > 0$ is the wave frequency, $\kappa$ is the local
epicyclic frequency of the disc, and $k$ is the radial wavenumber.
For a Keplerian disc, we have that $\kappa=\Omega$ and the radius
where $\omega = \kappa$ is an outer Lindblad resonance.  Its behaviour
in launching waves is essentially like that of an outer Lindblad
resonance encountered in non-axisymmetric cases (e.g.  Goldreich \&
Tremaine 1979).  The two-dimensional axisymmetric wave can propagate
only outside this radius (i.e. where $\omega > \Omega$) for a
Keplerian disc.

If such a wave behaves purely isothermally (i.e. $\gamma=1$), it will
propagate without loss to the outer radius of the disc.  However, as
we will see in this paper, adiabatic waves with $\gamma=5/3$ steepen
into shocks after propagating a finite distance. Shocks occur because
the sound speed at the crest of the wave is greater than that in the
trough.  Thus, a wave that is launched with a sinusoidal profile
eventually steepens into a saw-tooth form and its energy is dissipated
in a shock.  The number of wavelengths required for the shock
formation to occur can be $\sim c/\Delta c$, where $\Delta c$ is the
difference in sound speed between the crest and the trough of the
wave. This argument ignores various geometrical corrections that occur
in a multi-dimensional disc. In addition, it ignores the possibility
that dispersive effects could lessen the wave steepening (Larson 1990;
Papaloizou \& Lin 1995). This estimate suggests that the propagation
distance before shocks set in depends on the initial amplitude of the
wave, i.e. the wave amplitude at the resonance.

\begin{figure*}
\centerline{\psfig{figure=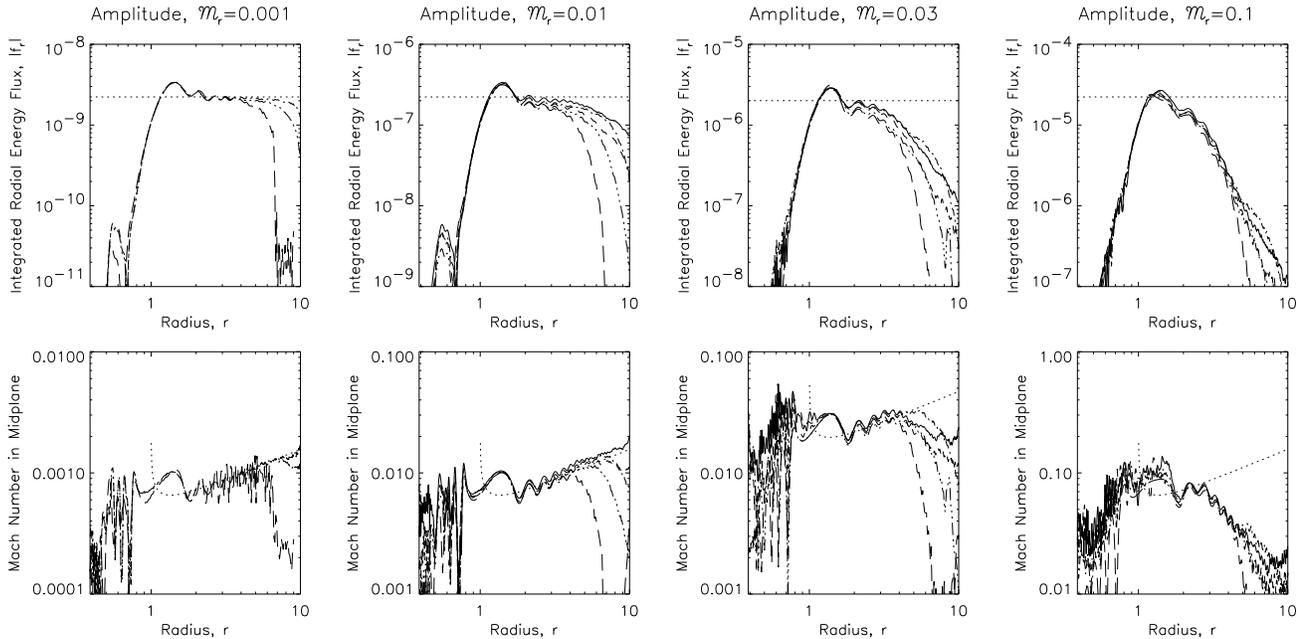,width=17.5truecm,height=8.75truecm,rwidth=17.5truecm,rheight=8.75truecm}}
\caption{\label{IF_ep0.10} The integrated radial energy flux, 
  $|f_{\rm r}|$, (top) and mean (time-averaged) Mach number at 
  the mid-plane of the disc, ${\cal M}(r, \theta=\pi/2)$,
  (bottom) are plotted as functions of radius for the \fe mode in a
  vertically isothermal disc with $H/r=0.1$.  The results are shown
  for four different wave amplitudes: ${\cal M}_{\rm r}=0.001$ (left),
  ${\cal M}_{\rm r}=0.01$ (centre-left), ${\cal M}_{\rm r}=0.03$
  (centre-right), ${\cal M}_{\rm r}=0.1$ (right).  Each case was
  performed with different numerical resolutions and/or viscosity to
  test for convergence: $250\times 111$ and $q_{\rm visc}=4$
  (long-dashed), $500\times 221$ and $q_{\rm visc}=4$
  (triple-dot-dashed), $500\times 221$ and $q_{\rm visc}=0.5$
  (short-dashed), $1000\times 441$ and $q_{\rm visc}=4$
  (dot-dashed), $1000\times 441$ and $q_{\rm visc}=0.5$ (solid).  The
  horizontal dotted line indicates 75 percent of the flux predicted by
  the Goldreich \& Tremaine formula (equation \ref{gt}) and the
  associated Mach number of the wave.  The value of 75 percent is that
  predicted by linear theory to be deposited in the \fe mode.  As can
  be seen from the figure, the hydrodynamic results are in excellent
  agreement with linear theory for the low-amplitude cases.  In the
  high-amplitude cases, the radial energy flux is attenuated by
  dissipation.}
\end{figure*}

\subsection{General wave properties}

\label{general}

Along with the simple two-dimensional wave, a vertically isothermal
disc admits series of other waves, all of which possess vertical
motion (Lubow \& Pringle 1993).  These wave modes can be categorised
by their vertical velocity structure. In a thermally stratified disc,
as pointed out by Lin, Papaloizou \& Savonije (1990), the
two-dimensional wave does not exist because $\gamma p/\rho$ is a
function of $z$ (cf.\ equation \ref{dispersion}). In such a disc, all
modes involve vertical motions, as follows from Korycansky \& Pringle
(1995).

In the spherical geometry of the simulations, $z=r\cos\theta$.  The
linear wave modes then have the form
\begin{equation}
  Q(r,\theta,t)={\rm Re}\left\{\tilde Q(r,\theta)\exp
  \left[-{\rm i}\omega t+{\rm i}\int k(r)\,{\rm d}r\right]\right\},
\end{equation}
where $Q$ is any perturbation quantity.  The frequency $\omega$ is
defined to be positive. The wavenumber $k(r)$ satisfies $|kr|\gg1$
except near resonances, where this WKB form breaks down. The
wavenumber can be positive or negative, depending on the radial
direction (inward or outward) of the phase velocity. The amplitude
$\tilde Q$ varies slowly with $r$.  The vertical structure of the
wave, and the relation between $\omega$ and $k$, are determined by an
eigenvalue problem, local in $r$, which admits a set of discrete
modes.  Each mode has a different dispersion relation
$\omega=\omega(k,r)$ which must be satisfied at every $r$.

Using the notation of Ogilvie (1998), these modes can be categorised
into four classes which are determined by the dominant forces involved
in their propagation.  The f (fundamental), p (pressure-dominated),
and g (buoyancy-dominated) modes propagate where $\omega>\Omega(r)$,
while r~modes (rotationally dominated modes) propagate where
$\omega<\Omega(r)$.  There are only two f~modes, \fe and \fo.  Each of
the other classes contains an infinite sequence of modes with
increasing numbers of nodes in the vertical structure.  These may be
classified as ${\rm p}_1^{\rm e}$, ${\rm p}_2^{\rm e}$, ${\rm
  p}_3^{\rm e}$, \dots, ${\rm p}_1^{\rm o}$, ${\rm p}_2^{\rm o}$,
${\rm p}_3^{\rm o}$, \dots, etc., where the number refers to the order
of the mode, and `e' or `o' refers to even or odd symmetry about the
mid-plane.

In this paper, we focus our attention on the following modes.
\begin{itemize}
\item the \fe mode, which propagates where $\omega > \Omega(r)$.  In a
  vertically isothermal disc, this mode is the two-dimensional mode
  described in the previous subsection.  For other types of discs, the
  \fe mode behaves like the two-dimensional wave close to the
  resonance, but behaves like a surface gravity wave away from the
  resonance (Ogilvie 1998; Lubow \& Ogilvie 1998).
\item the \pe mode, which propagates where $
  \omega>\Omega(r)\sqrt{\gamma+1}$.  Its vertical velocity at
  resonance is proportional to $z$ (or $\cos\theta$).
\item the \foro mode, which propagates at all radii, but has a
  resonance where $\omega=\Omega$.  Its vertical velocity
  eigenfunction at resonance is independent of $z$, while its radial
  velocity is proportional to $z$.  The \foro mode is a corrugation
  wave or disc warp, where the mid-plane of the disc oscillates up and
  down.
\end{itemize}

To excite these modes efficiently, we apply an acceleration $(a_r,
a_{\theta})$ (force per unit mass) to our equilibrium discs of the
following forms.
\begin{enumerate}
\item $a_r={\cal A}(r)\sin\omega t$ and $a_{\theta}=0$ for the \fe
  mode, with ${\cal A}(r)={\cal A}_0r^{-2}$;
  \item $a_r=0$ and $a_\theta={\cal A}(r)r\cos\theta\sin\omega t$
    for the \pe mode, with ${\cal A}(r)={\cal A}_0r^{-3}$;
  \item $a_r=0$ and $a_\theta={\cal A}(r)\sin\omega t$ for the
    \foro mode, with ${\cal A}(r)={\cal A}_0r^{-2}$.
\end{enumerate}
In each case, the power-law dependence of ${\cal A}$ on $r$ is chosen
so that the amplitude of the non-resonant response can be reasonably
small (linear) at both small and large radii. We choose a wave
frequency $\omega=1$ so that the resonance in our dimensionless units
($\omega=\Omega$ or $\omega=\Omega\sqrt{\gamma+1}$) lies at $r=1$ (for
the \fe and corrugation modes) or $r=(\gamma+1)^{1/3}\approx 1.39$
(for the \pe mode).  The simulated disc contains these resonance
locations.  The acceleration is applied from the beginning of the
calculations and its strength is parameterised by ${\cal A}_0$.

Each form of the acceleration listed above generally results in the
excitation of a variety of modes.  The total energy flux generated at
the resonance that is carried by all modes for an axisymmetric disc
can be determined by adapting the torque formula of Goldreich \&
Tremaine (1979).  In Appendix B, we derive the total energy flux for
each form of acceleration listed above.

In practice, it is more convenient to express the strength of the
forcing in terms of the spatial peak of the time-averaged wave Mach
number at the disc mid-plane that occurs near the resonance, which we
denote as ${\cal M}_{\rm r}$. In terms of our notation, we have
\begin{equation}
\label{mc}
  {\cal M}_{\rm r} = \max[{\cal M}(r, \theta = \pi/2)],
\end{equation}
where ${\cal M}$ is defined by equation (\ref{macheq}) and the maximum
is taken over a suitable neighbourhood of the resonance ($0.8 < r <
1.2$).  This measure of forcing strength is used in comparing the
simulations involving different disc models. In carrying out a
simulation with a chosen value of ${\cal M}_{\rm r}$, we vary ${\cal
  A}_0$ until the desired value of ${\cal M}_{\rm r}$ is achieved.

\begin{figure}
\centerline{\psfig{figure=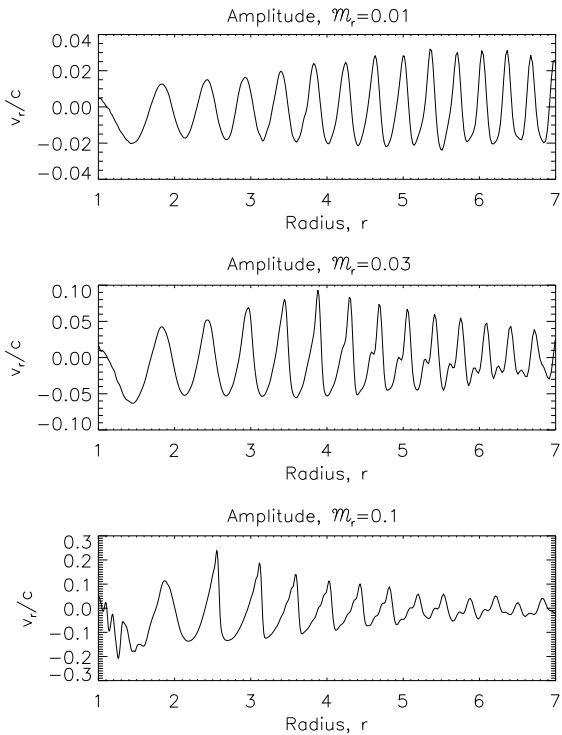,width=8.0truecm,height=10.0truecm,rwidth=8.0truecm,rheight=10.0truecm}}
\caption{\label{IF_ep0.10_peak} The Mach number of the radial motion
  of the wave in the mid-plane of the disc is plotted at a particular
  instant as a function of radius for the f$^{\rm e}$ mode in the
  vertically isothermal disc with $H/r=0.1$.  Various wave amplitudes
  are shown: ${\cal M}_{\rm r}=0.01$ (top), ${\cal M}_{\rm r}=0.03$
  (middle) and ${\cal M}_{\rm r}=0.1$ (bottom).  For the low-amplitude
  case, ${\cal M}_{\rm r}=0.01$, the wave maintains its sinusoidal
  profile until numerical dissipation occurs.  For the higher
  amplitudes, ${\cal M}_{\rm r}=0.03$ and ${\cal M}_{\rm r}=0.1$, the
  wave steepens, shocks and dissipates.  This indicates that the
  physical dissipation is numerically resolved.}
\end{figure}

\subsection{Numerical convergence}

\label{convergence}

As described above, in a vertically isothermal disc, the \fe mode
should propagate outwards from the resonance without a loss of flux
until the wave steepens to form shocks.  One of the main difficulties
in obtaining non-linear results is spatially resolving the wave until
this wave steepening and shock formation occur.  If the flow becomes
unresolved (i.e. it varies on a scale similar to the zone spacing),
the wave will dissipate energy.  This dissipation occurs in one of two
ways.  In the absence of an artificial viscosity, the energy carried
by the wave is lost from the calculation.  If there is an artificial
viscosity, as is employed in the calculations presented here, the wave
energy is converted into thermal energy.  Artificial viscosity
provides a means of simulating shocks (flow discontinuities) in the
code.  Numerical dissipation in the ZEUS-2D code is proportional to
the square of the velocity difference across a grid zone.  Thus, for
an approximately linear wave, the numerical dissipation is
proportional to $q_{\rm visc} A^2 /(\lambda N_r)$, where $A$ and
$\lambda$ are the amplitude and wavelength of the wave, respectively.

There are several ways to determine whether a wave dissipates
numerically or physically.  First, the resolution of the calculations
can be increased.  If the dissipation is numerical, the wave will
propagate further before it dissipates.  Secondly, keeping the
resolution fixed, the strength of the artificial viscosity, $q_{\rm
  visc}$, can be varied.  If the dissipation is numerical, the wave
will propagate further before it dissipates with lower $q_{\rm visc}$.
Note, however, that if $q_{\rm visc}\ll 1$, energy will simply be lost
from the calculation instead of being converted to thermal energy
(e.g. calculations with $q_{\rm visc}= 0.1$ and $q_{\rm visc}= 0.01$
give almost identical results).  The third way is to examine the form
of the wave to see if wave steepening occurs just before the wave
dissipates.  If the wave maintains a sinusoidal profile, the
dissipation is numerical.  In order to determine which of the
non-linear results are due to physical dissipation, we use all three
of these indicators.

\section{The vertically isothermal disc}

In this section, we study the excitation, propagation and dissipation
of adiabatic ($\gamma = 5/3$) waves in a vertically isothermal
accretion disc.  First, we consider the \fe mode (a plane wave).
Subsequently, we discuss the \pe mode and the \foro (corrugation)
mode.

\subsection{\fe mode}
\label{femode}

\subsubsection{Linear theory}

As discussed in Section \ref{excitation}, we adopt a purely radial
driving force per unit mass of $a_r(r, t) = {\cal A}(r) \sin\omega t$,
in order to efficiently excite the \fe mode.  The \fe mode corresponds
to the two-dimensional wave in a vertically isothermal disc.  Linear
theory predicts that the radial forcing of Section \ref{excitation}
should lead to the outwardly propagating \fe mode receiving 74.5\% of
the total radial energy flux (see Appendix B1).  The inwardly
propagating \re mode should receive about 6.0\% of the total flux. The
remainder is goes into g~modes (see Appendix B1).

According to linear theory (Lubow \& Pringle 1993), the \fe mode
should occupy the entire vertical thickness of the disc and propagate
outwards in radius.  The mean Mach number at the mid-plane of the disc
should increase as ${\cal M}(r, \theta=\pi/2) \propto r^{1/2}$ for $r
\gg 1$.  This weak dependence of the Mach number on $r$ means that a
low-amplitude wave is expected to propagate a large distance before it
steepens into a shock and dissipates.  The vertical structure of the
wave can also be predicted by linear theory.  The mean Mach number
varies vertically as ${\cal M} \propto \exp(z^2/5H^2)$, for
$\gamma=5/3$.  Thus, although a wave may be linear near the mid-plane,
there will be some finite height at which the motions are supersonic
and we would expect shocks to form.  However, the majority of the wave
energy is contained near the mid-plane and therefore this does not
diminish the wave flux appreciably.

The r~modes will be discussed in detail in Section \ref{foromode}.
Briefly, however, the \re mode should propagate inwards and be
channelled towards the mid-plane of the disc.

\subsubsection{Low-amplitude, non-linear results}

We turn now to the non-linear hydrodynamic calculations performed with
ZEUS-2D.  Figure \ref{IF_ep0.10} plots the magnitude of the vertically
integrated radial energy flux, \fr (equation \ref{fluxeq}), and the
mean Mach number (equation \ref{macheq}) in the mid-plane of the disc,
${\cal M}(r, \theta=\pi/2)$, as functions of $r$.  They are plotted
for four values of the Mach number near resonance ${\cal M}_{\rm r}$.
Notice that the radial energy flux is in fact negative for $r<0.7$.
This effect is due the inwardly propagating \re mode that is excited
along with the \fe mode.

The dotted lines in the upper panels of Figure \ref{IF_ep0.10} show
the contribution from the \fe mode to the total radial energy flux
that is predicted by linear theory.  In the lower panels, we plot the
corresponding Mach numbers that are predicted in the mid-plane by
linear theory.  The non-linear results are in excellent agreement with
the linear theory.

Consider the weakest \fe mode (left panels, Figure \ref{IF_ep0.10}).
Excitation of the \fe mode occurs over a finite radial extent (a few
$H$) around $r=1$.  For $r > 1$, the flux reaches a peak value just
beyond the resonance.  The drop in flux just beyond the peak is
probably due to the rapid dissipation of g~modes, which account for
about 20\% of the total flux.  The \fe mode continues to propagate to
greater radii and its flux remains nearly constant for $ 2< r <8$. The
value of the flux there agrees well with the predicted \fe mode flux.
The dissipation that occurs at large radii is numerical as seen by the
lack of convergence with increasing resolution.  Thus, the wave should
continue to propagate outwards, beyond the grid boundary.

For $r <1$, the \re mode propagates inwards and overlaps with the
evanescent tail of the \fe mode.  The \re mode is only modelled for $r
\gsim 0.5$ because the edge of the equilibrium disc is at $R_1=0.5$.
The two modes have fluxes of opposite sign, and the net flux changes
sign near $r = 0.7$.  The peak radial energy flux from the simulation
in this region is about 3\% of the total flux. This result is
consistent with the expected \re mode negative flux of about 6\% of
the total overlapping with the tail of the positive \fe mode flux.
The r~modes will be discussed in more detail in section \ref{foromode}
where an \ro mode receives 50\% of the total flux.

These results are reinforced by Figure \ref{IF_ep0.10_cont}, which
gives contour plots of the wave energy density (equation \ref{energyeq}) 
and the energy dissipation rate for the low-amplitude calculations
(left panels).  
There is a lot of dissipation at high altitude in the disc in the range
$1<r<2$, as expected for the g~modes.  In the bulk of the disc, there
is no significant dissipation and the \fe mode propagates outwards
with virtually no loss of flux (a small amount is lost at high
altitudes where the motions become supersonic).

The behaviour of the inwardly propagating \re mode can also be seen in
the left-hand panels of Figure \ref{IF_ep0.10_cont}.  As predicted by
linear theory (Lubow \& Pringle 1993), the wave is channelled towards
the mid-plane.  This will be discussed further in Section
\ref{foromode}.

\begin{figure*}
\vspace{17.5truecm}
\caption{\label{IF_ep0.10_cont} The case considered here is the \fe mode
  in a vertically isothermal disc with $H/r=0.1$.  The kinetic energy
  density in the wave ($\log E$, equation \ref{energyeq}) is shown in
  the upper panels.  The energy dissipation rate per unit volume is
  shown in the lower panels as a logarithmic grey-scale.  The grey-scales
  cover three orders of magnitude.  The left
  panels are for amplitude ${\cal M}_{\rm r}=0.001$ and the right
  panels are for amplitude ${\cal M}_{\rm r}=0.1$.  At low amplitude
  the \fe mode propagates to large radius throughout the body of the
  disc unaffected by the small amount of dissipation that occurs at
  high $|z|$.  In the high-amplitude case dissipation occurs in the
  body of the disc and significantly attenuates the wave (see Figure
  \ref{IF_ep0.10}).
  In both cases the grid resolution is $N_r\times N_{\theta}=1000 \times 441$
  and $q_{\rm visc}=4.0$.}
\end{figure*}

\begin{figure*}
\centerline{\psfig{figure=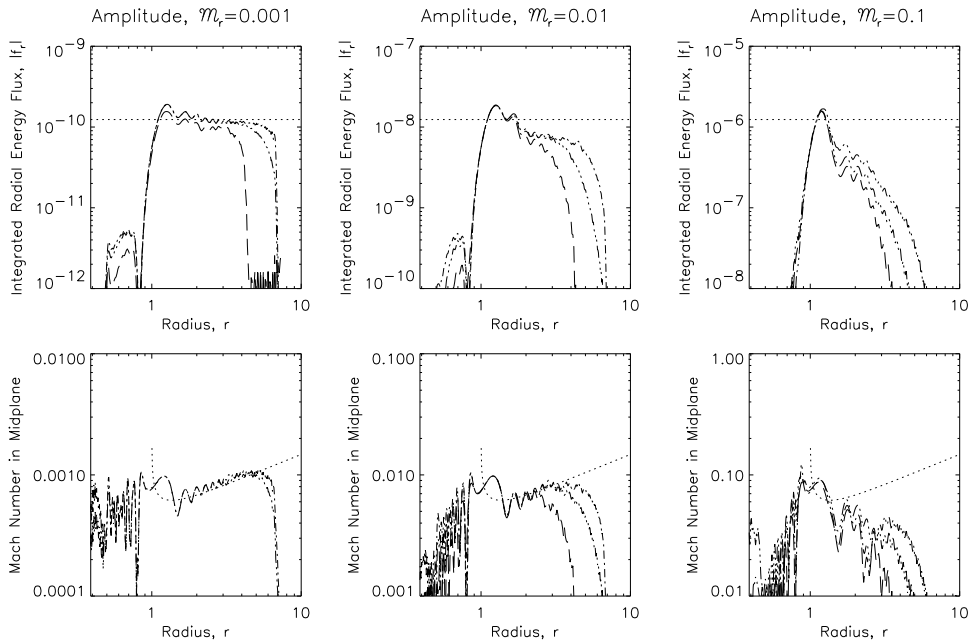,width=13.125truecm,height=8.75truecm,rwidth=13.125truecm,rheight=8.75truecm}}
\caption{\label{IF_ep0.05}  The integrated radial energy flux,
  $|f_{\rm r}|$, (top) and 
  mean (time-averaged) Mach number at the mid-plane of the disc, 
  ${\cal M}(r, \theta=\pi/2)$,
  (bottom) are plotted as functions of radius for the \fe mode in a
  vertically isothermal disc with $H/r=0.05$.  The results are shown
  for three different wave amplitudes: ${\cal M}_{\rm r}=0.001$
  (left), ${\cal M}_{\rm r}=0.01$ (centre), ${\cal M}_{\rm r}=0.1$
  (right).  Each case was performed with different numerical
  resolutions to test for convergence: $250\times 111$ and $q_{\rm
    visc}=4$ (long-dashed), $500\times 221$ and $q_{\rm visc}=4$
  (triple-dot-dashed), $1000\times 441$ and $q_{\rm visc}=4$
  (dot-dashed).  The horizontal dotted line indicates 75 percent of
  the flux predicted by the Goldreich \& Tremaine formula (equation
  \ref{gt})and the associated Mach number of the wave.  The value of
  75 percent is that predicted by linear theory to be deposited in the
  \fe mode.  As can be seen from the figure, the hydrodynamic results
  are in excellent agreement with linear theory for the lowest
  amplitude case.  In the high-amplitude cases, the radial energy flux
  is attenuated by dissipation.}
\end{figure*}

\subsubsection{Dependence on wave amplitude}

As the driving amplitude is increased, both numerical and physical
dissipation occur more quickly.  Numerical dissipation increases as
the square of the amplitude of the wave (Section \ref{convergence}).
Thus, waves with larger values of ${\cal M}_{\rm r}$ propagate a
shorter distance before numerical dissipation occurs (cf.\ Figure
\ref{IF_ep0.10}, results with ${\cal M}_{\rm r}=0.001$, and ${\cal
  M}_{\rm r}=0.01$).

Physical dissipation occurs if the sinusoidal wave steepens into a
shock, which takes place in $\sim c/\Delta c$ wavelengths from the
resonance (Section \ref{twodimensional}).  For waves with very small
values of ${\cal M}_{\rm r}$, steepening requires many wavelengths and
numerical dissipation sets in first, as we saw in the case for ${\cal
  M}_{\rm r}=0.001$.  Physical dissipation in our simulations was
obtained for cases with ${\cal M}_{\rm r} = 0.03$ and $0.1$.  Consider
the panels for ${\cal M}_{\rm r} = 0.03$ in Figure \ref{IF_ep0.10}.
Convergence of flux and ${\cal M}(r,\theta=\pi/2)$ is obtained for
$r\lsim 4$ with resolutions of $500\times 221$ and $1000\times 441$.
With an initial amplitude of ${\cal M}_{\rm r} = 0.1$, even a
resolution of $250\times 111$ is sufficient to obtain physical
dissipation.  In the ${\cal M}_{\rm r} = 0.03$ case, the wave flux has
been reduced by an order of magnitude by $r\approx 10$, whereas in the ${\cal
  M}_{\rm r} = 0.1$ case this level of reduction occurs at $r\approx
4$.

To emphasise that this dissipation is physical rather than numerical,
we plot in Figure \ref{IF_ep0.10_peak} the radial velocity profile of
the wave in the mid-plane for the cases with ${\cal M}_{\rm r} =0.01,
0.03$, and $0.1$.  Steepening of the wave towards the profile of a
shock-wave is clearly visible in those cases with ${\cal M}_{\rm r}
=0.03$ and 0.1, whereas in the ${\cal M}_{\rm r} = 0.01$ calculation
the profile remains sinusoidal indicating only numerical, rather than
physical, dissipation.  This result is consistent with Figure
\ref{IF_ep0.10_cont} which shows that increasing ${\cal M}_{\rm r}$
decreases the volume in which the \fe mode propagates.

\subsubsection{Dependence on disc thickness}

\begin{figure}
\centerline{\psfig{figure=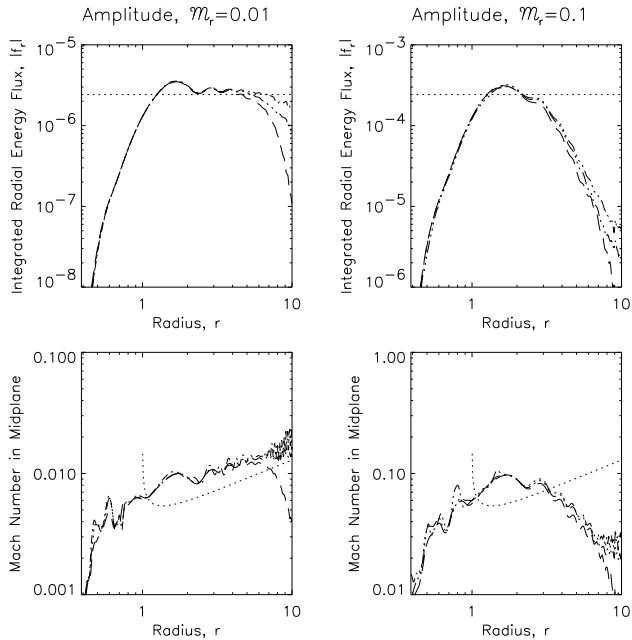,width=8.75truecm,height=8.75truecm,rwidth=8.75truecm,rheight=8.75truecm}}
\caption{\label{IF_ep0.20} The integrated radial energy flux,
  $|f_{\rm r}|$, (top) and 
  mean (time-averaged) Mach number at the mid-plane of the disc, 
  ${\cal M}(r, \theta=\pi/2)$,
  (bottom) are plotted as functions of radius for the \fe mode in a
  vertically isothermal disc with $H/r=0.2$.  The results are shown
  for two different wave amplitudes: ${\cal M}_{\rm r}=0.01$ (left),
  ${\cal M}_{\rm r}=0.1$ (right).  Each case was performed with
  different numerical resolutions to test for convergence: $250\times
  111$ and $q_{\rm visc}=4$ (long-dashed), $500\times 221$ and $q_{\rm
    visc}=4$ (triple-dot-dashed), $1000\times 441$ and $q_{\rm
    visc}=4$ (dot-dashed).  The horizontal dotted line indicates 75
  percent of the flux predicted by the Goldreich \& Tremaine formula
  (equation \ref{gt}) and the associated Mach number of the wave.  The
  value of 75 percent is that predicted by linear theory to be
  deposited in the \fe mode.  As can be seen from the figure, the
  hydrodynamic results for the energy flux are in good agreement with
  linear theory for the lowest amplitude case, but for this relatively
  thick disc the predicted Mach number underestimates the computed
  value by about 30 percent.  In the high-amplitude case, the radial
  energy flux is attenuated by dissipation. }
\end{figure}

In Figures \ref{IF_ep0.05} and \ref{IF_ep0.20}, we plot the energy
flux and mean Mach number ${\cal M}(r, \theta = \pi/2)$ as functions
of $r$ for discs with $H/r=0.05$ and $H/r=0.2$, respectively.  The
main difference between the waves in these discs and those in the
$H/r=0.1$ disc is that the wavelength is proportional to $H/r$.
Since the numerical dissipation is $\propto 1/(\lambda N_r)$ (Section
\ref{convergence}), a wave in a thick disc can be followed to larger
radii than in a thin disc for the same radial resolution (cf.  ${\cal
  M}_{\rm r}=0.01$ in Figures \ref{IF_ep0.10}, \ref{IF_ep0.05} and
\ref{IF_ep0.20}).  Thus, for example, waves with ${\cal M}_{\rm r}=0.1$ 
are well resolved in the hottest disc, $H/r=0.2$, even with
$N_r\times N_{\theta}=250\times 111$, and dissipate due to shocks,
whereas in the coldest disc, $H/r=0.05$, even the highest resolutions
do not give convincing convergence (numerical dissipation still plays
a significant role).

\begin{figure*}
\centerline{\psfig{figure=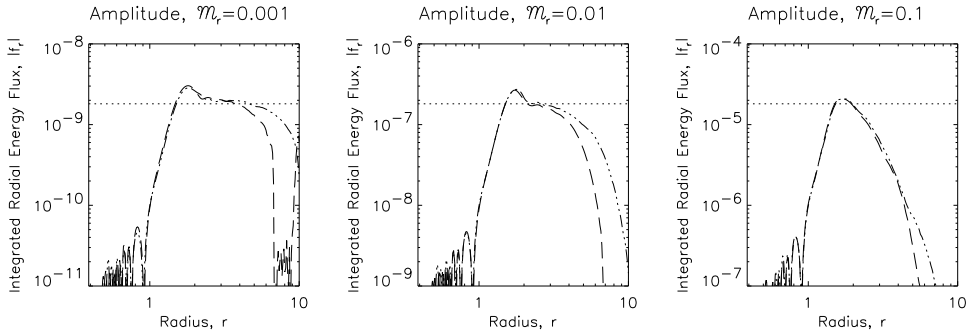,width=13.125truecm,height=4.362truecm,rwidth=13.125truecm,rheight=4.362truecm}}
\caption{\label{IP_ep0.10}  The integrated radial energy flux,
  $|f_{\rm r}|$,  is 
  plotted as a function of radius for the \pe mode in a vertically
  isothermal disc with $H/r=0.1$.  The results are shown for three
  different wave amplitudes: ${\cal M}_{\rm r}=0.001$ (left), ${\cal
    M}_{\rm r}=0.01$ (centre), ${\cal M}_{\rm r}=0.1$ (right).  Each
  case was performed with different numerical resolutions to test for
  convergence: $250\times 111$ and $q_{\rm visc}=4$ (long-dashed),
  $500\times 221$ and $q_{\rm visc}=4$ (triple-dot-dashed).  The
  horizontal dotted line indicates the flux predicted by linear theory
  (equation \ref{linear_flux_p}).  As can be seen from the figure, the
  hydrodynamic results are in excellent agreement with linear theory
  for the low amplitude cases.  At high amplitude, the radial energy
  flux is attenuated by dissipation.}
\end{figure*}

Strictly, linear theory assumes $H \ll r$.  However, we find that,
even with $H/r=0.2$, linear theory provides a good description of the
wave excitation and propagation.  A further prediction that can be
derived from linear theory is that the ratio of the Mach number at the
resonance to the Mach number at $r \gg r_{\rm res}$ should be nearly
independent of $H/r$.  This is confirmed by the non-linear
calculations presented here.

\subsection{p$^{\rm e}_{1}$ mode}

\label{pemode}

\subsubsection{Linear theory}

As discussed in Section \ref{excitation}, we adopt a driving force per
unit mass that is purely in the $\theta$-direction of the form
$a_\theta(r, \theta, t) = {\cal A}(r) r\cos\theta\sin\omega t$, in
order to efficiently excite the \pe mode.  The \pe mode is launched at
a vertical resonance $r \approx 1.39$ and propagates outwards.  We
predict its radial energy flux in Appendix B2.  No
other wave is expected.

Unlike the \fe mode, the \pe mode has vertical structure (see Lubow \&
Pringle 1993).  In the vicinity of the resonance, the wave energy has
a minimum on the mid-plane of the disc and two maxima at $z/H \approx
2$.  This is due to the form of the excitation which depends linearly
on $z$ (Section \ref{excitation}).  For $r \gsim 1.7$, this changes to
three maxima, one on the mid-plane and two at $z/H \approx 3$, with two
minima at $z/H \approx 2$.  Lubow \& Pringle (1993) interpreted this
as the wave energy rising up within the disc as the wave propagates
outwards.  In fact, however, the vertical distribution of the wave
energy does not change beyond $r\approx 2$.  The wave propagates
outwards maintaining the above distribution of energy indefinitely.
As with the \fe mode, the wave is eventually expected to undergo
steepening into a shock and dissipate, but a low-amplitude wave will
propagate for a large distance before this takes place.

\subsubsection{Low-amplitude, non-linear results}

Figure \ref{IP_ep0.10} plots the magnitude of the vertically
integrated radial energy flux, \fr (equation \ref{fluxeq})
as a function of $r$.  We consider three values of the mean Mach 
number near resonance ${\cal M}_{\rm r}$.  As with the case for 
the \fe mode, the low-amplitude \pe mode
propagates outwards from the resonance until most of the energy is
dissipated numerically.

A small negative energy flux is seen for $r \lsim 0.9$.  We attribute
this to a low-amplitude r~mode that is excited at the Lindblad
resonance at $r=1$.  In a very thin disc, vertical forcing of the type
applied here should not excite any modes at the Lindblad resonance.
However, when $H/r=0.1$, there is some ambiguity between the meanings
of `horizontal' and `vertical' away from the mid-plane.  As a result,
some launching at the Lindblad resonance is possible at the level of
$(H/r)^2$.

In Figure \ref{IP_ep0.10_cont} we plot the wave energy and energy
dissipation rate for the low-amplitude \pe mode (left panels).  
The vertical distribution of the wave energy density is in excellent 
agreement with linear theory in the bulk of the disc with only a small
amount of dissipation at high altitude where the motions become sonic.

\begin{figure*}
\vspace{17.5truecm}
\caption{\label{IP_ep0.10_cont} The case considered here is the \pe mode
  in a vertically isothermal disc with $H/r=0.1$.  The kinetic energy
  density in the wave ($\log E$, equation \ref{energyeq}) is shown in
  the upper panels covering four orders of magnitude.  The energy 
  dissipation rate per unit volume is shown in the lower panels as a 
  logarithmic grey-scale covering six orders of magnitude.  The left
  panels are for amplitude ${\cal M}_{\rm r}=0.001$ and the right
  panels are for amplitude ${\cal M}_{\rm r}=0.1$.  At low amplitude
  the \pe mode propagates to large radius throughout the body of the
  disc unaffected by the small amount of dissipation that occurs at
  high $|z|$.  The \pe mode displays the vertical structure predicted
  by linear theory.  In the high-amplitude case dissipation occurs in
  the body of the disc and significantly attenuates the wave (see
  Figure \ref{IP_ep0.10}).
  In both cases the grid resolution is $N_r\times N_{\theta}=500 \times 221$
  and $q_{\rm visc}=4.0$.}
\end{figure*}

\subsubsection{Wave amplitude and disc thickness}

As with the \fe mode, the
dissipation occurs more rapidly with increased driving.  Calculations
with ${\cal M}_{\rm r}=0.1$ resolve the wave until physical
dissipation due to wave steepening occurs, but the lower ${\cal
  M}_{\rm r}$ cases are still limited by numerical dissipation.

We have not performed calculations of the \pe mode for discs with
different thicknesses since, as we found for \fe mode, we do not
expect any significant dependence of the propagation of the \pe mode
on $H/r$.  As with the \fe mode, the longer wavelength in hotter discs
would make it easier to resolve the wave to larger radii.

\begin{figure*}
\centerline{\psfig{figure=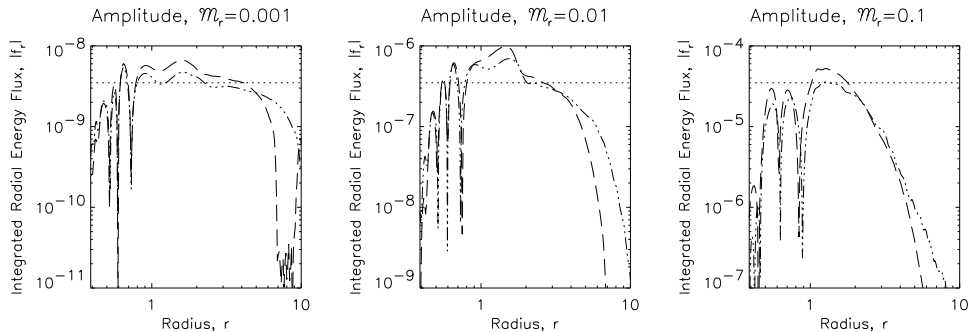,width=13.125truecm,height=4.362truecm,rwidth=13.125truecm,rheight=4.362truecm}}
\caption{\label{IT_ep0.10}  The integrated radial energy flux,
  $|f_{\rm r}|$,  is 
  plotted as a function of radius for the \foro (corrugation) mode in
  a vertically isothermal disc with $H/r=0.1$.  The results are shown
  for three different wave amplitudes: ${\cal M}_{\rm r}=0.001$
  (left), ${\cal M}_{\rm r}=0.01$ (centre), ${\cal M}_{\rm r}=0.1$
  (right).  Each case was performed with different numerical
  resolutions to test for convergence: $250\times 111$ and $q_{\rm
    visc}=4$ (long-dashed), $500\times 221$ and $q_{\rm visc}=4$
  (triple-dot-dashed).  The horizontal dotted line indicates the flux
  predicted by linear theory (equation \ref{linear_flux_foro}).  For
  this case, the inwardly propagating \ro mode ($r<1$) and the
  outwardly propagating \fo mode ($r>1$) carry equal amounts of
  energy.  As can be seen from the figure, the hydrodynamic results
  are in reasonable agreement with linear theory for the low-amplitude
  high-resolution case.  At high amplitude, the radial energy flux is
  attenuated by dissipation.}
\end{figure*}

\begin{figure*}
\vspace{17.5truecm}
\caption{\label{IT_ep0.10_cont} The case considered here is the \foro 
  (corrugation) mode in a vertically isothermal disc with $H/r=0.1$.
  The kinetic energy density in the wave ($\log E$, equation
  \ref{energyeq}) is shown in the upper panels covering five orders of 
  magnitude.  The energy
  dissipation rate per unit volume is shown in the lower panels as a
  logarithmic grey-scale covering six orders of magnitude.  
  The left panels are for amplitude ${\cal
    M}_{\rm r}=0.001$ and the right panels are for amplitude ${\cal
    M}_{\rm r}=0.1$.  At low amplitude the \fo mode propagates
  outwards from $r=1$ to large radius throughout the body of the disc
  unaffected by the small amount of dissipation that occurs at high
  $|z|$.  The \ro mode propagates inwards from $r=1$ and is channelled
  towards the mid-plane of the disc.  Both the \fo mode and the \ro
  mode display the vertical structure predicted by linear theory.  In
  the high-amplitude case dissipation occurs in the body of the disc
  and significantly attenuates the wave (see Figure \ref{IT_ep0.10}).
  In both cases the grid resolution is $N_r\times N_{\theta}=500 \times 221$
  and $q_{\rm visc}=4.0$.}
\end{figure*}

\subsection{f$^{\rm o}$/r$^{\rm o}_{1}$ mode}

\label{foromode}

\subsubsection{Linear theory}

As discussed in Section \ref{excitation}, we adopt a driving force per
unit mass that purely in the $\theta$-direction of the form
$a_\theta(r, t) = {\cal A}(r) \sin\omega t$, in order to efficiently
excite the \foro mode.  The \foro mode, or corrugation mode, is also
launched at a vertical resonance although, in a Keplerian disc, this
coincides with the Lindblad resonance making it a hybrid
vertical/Lindblad resonance.  In the discs studied here, this hybrid
resonance occurs at $r=1$.  The \foro mode is able to propagate
throughout the entire disc with equal amounts of flux in the inwardly
propagating \ro mode and the outwardly propagating \fo mode.  In
Appendix B3, we determine the total energy flux produced at this
resonance.

The \ro mode is rapidly channelled towards the mid-plane of the disc
with an associated rapid rise in the Mach number of the wave that
might be expected to lead to dissipation of the wave in shocks (Lubow
\& Pringle 1993).  The \fo mode has two peaks in the vertical
distribution of its wave energy with a minimum on the disc mid-plane.

\subsubsection{Low-amplitude, non-linear results}

Figure \ref{IT_ep0.10} plots the magnitude of the vertically
integrated radial energy flux, \fr (equation \ref{fluxeq})
as a function of $r$.  We consider three values of the mean 
Mach number near resonance ${\cal M}_{\rm r}$.  
As predicted by linear theory, this mode propagates throughout
the entire disc.  Outward of the resonance ($r>1$), the wave
propagates as an \fo mode.  Inward of the resonance, the wave
propagates as an \ro mode.  An equal amount of flux goes into each of
the modes, and the magnitude of the flux is in good agreement with
linear theory, particularly for the calculations with the highest 
resolution.

Once again, for low-amplitude driving, the distance that the waves
propagate is limited by numerical dissipation rather than physical
dissipation (shocks).

The energy of the \fo mode is concentrated away from the mid-plane of
the disc (Figure \ref{IT_ep0.10_cont}) and its vertical distribution
is in excellent agreement with linear theory for $|z/H| \lsim 3$.
Above this region and particularly 
within the range $1 \le r \lsim 2.5$, shocks
form and a small amount of energy is dissipated.  However, the
majority of the wave energy remains within $|z/H| \lsim 3$ and
propagates outwards indefinitely.  Lubow \& Pringle (1993) noted the
rise of wave energy away from the mid-plane that occurs from near the
resonance to $r=3$ and interpreted this as the wave energy rising into
the atmosphere.  However, we see here that this `rising' does not
continue indefinitely; the wave simply occupies the bulk of the disc
as it propagates to large radii, in a similar manner to the \fe mode
except that it has vertical structure.

As expected from linear theory (Lubow \& Pringle 1993), the inwardly
propagating \ro mode is strongly focused towards the mid-plane of the
disc (Figure \ref{IT_ep0.10_cont}).  Both linear theory and the
non-linear results also show that the wavelength of the \ro mode
decreases towards the centre.

\subsubsection{Wave amplitude}

With strong driving (e.g. ${\cal M}_{\rm r}=0.1$), physical
dissipation through shocks occurs in the simulations and those
with different resolution converge, at least for the \fo mode.  By the
time the \fo mode reaches $r=10$, the flux has been reduced by a
factor of $\sim 10^3$.  The wave is confined to a region very near the
mid-plane of the disc; shocks are present at large $|z|$ (Figure
\ref{IT_ep0.10_cont}, right panels).  We have not performed
calculations of the corrugation mode with other disc thicknesses.

\section{The polytropic disc}
\label{secpoly}

We now turn our attention to polytropic discs.  The disc studied in
this section has an optical depth $\tau=25000$, and only a tenuous
isothermal atmosphere.  Discs that have $\tau = 5-100$ (i.e.
intermediate between a vertically isothermal disc and a very optically
thick disc) are briefly described in Section 7.

\subsection{f$^{\rm e}$ mode}

\subsubsection{Linear theory}

As discussed in Section \ref{excitation}, we adopt a purely radial
driving force per unit mass of $a_r(r, t) = {\cal A}(r)\sin\omega t$,
in order to efficiently excite the \fe mode.  The \fe mode corresponds
to the two-dimensional wave in a vertically isothermal disc.  Linear
theory predicts that this radial forcing 
should lead to the outwardly propagating \fe mode receiving 98.3\% of
the total radial energy flux (see Appendix B1).  A greater fraction
goes into the \fe mode in the polytropic disc than in the vertically
isothermal disc.  The remainder goes into an inwardly propagating \re
mode.

The linear behaviour of the \fe mode in a polytropic disc has been
studied by Korycansky \& Pringle (1995), Lubow \& Ogilvie (1998) and
Ogilvie \& Lubow (1999).  Although the \fe mode is excited throughout
the entire vertical extent of the disc at the resonance, it is
channelled towards the surface of the polytropic disc, unlike in the
vertically isothermal disc.  The rate of channelling (the distance
from the resonance at which the wave energy becomes concentrated away
from the disc mid-plane) depends only on the degree of vertical
stratification, which is determined by the index of the polytrope, and
the azimuthal wavenumber $m$, which is fixed at $m=0$ for the
axisymmetric case. The rate is independent of the disc thickness $H$,
in the limit that $H \ll r$.  This concentration of the wave energy
into a small volume of cool, low-density gas near the disc surface (or
atmosphere) leads to a rapid increase of the Mach number of the wave with
radius.  Thus we expect that waves in polytropic discs will undergo
shocks more quickly (closer to the resonance) than waves in the
isothermal case described earlier.  Lubow \& Ogilvie (1998) speculated
that once the Mach number of the wave became of order unity, the wave
would shock and thus, such waves would not propagate over large
distances.  One of the aims of the non-linear hydrodynamic
calculations is to determine where this dissipation occurs.

\subsubsection{Low-amplitude, non-linear results}

\begin{figure*}
\centerline{\psfig{figure=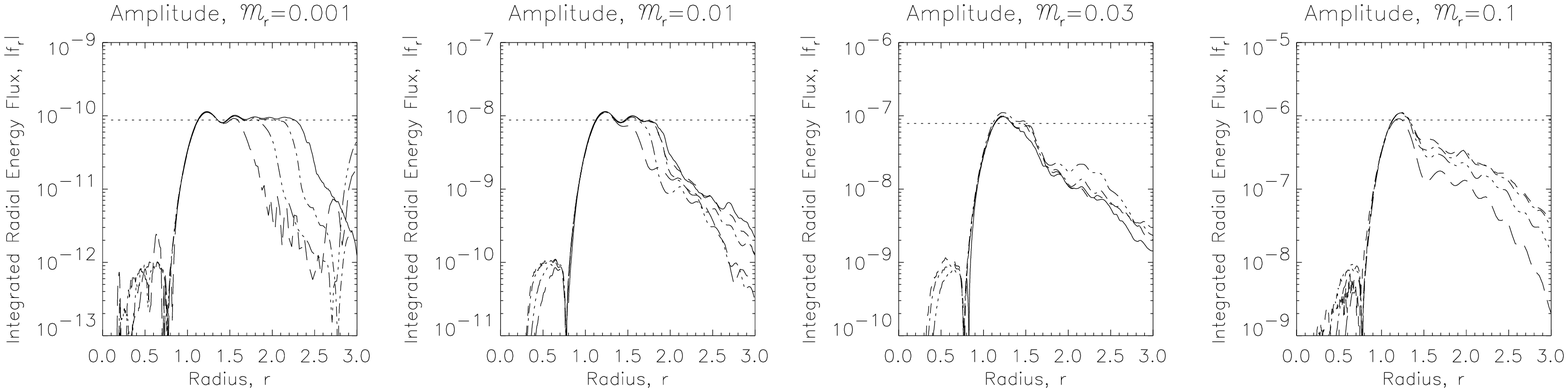,width=17.5truecm,height=4.362truecm,rwidth=17.5truecm,rheight=4.362truecm}}
\caption{\label{PF_ep0.10}  The integrated radial energy flux,
  $|f_{\rm r}|$,  is plotted 
  as a function of radius for the \fe mode in a polytropic disc with
  $H/r=0.1$ and $\tau=25000$.  The results are shown for four
  different wave amplitudes: ${\cal M}_{\rm r}=0.001$ (left), ${\cal
    M}_{\rm r}=0.01$ (centre-left), ${\cal M}_{\rm r}=0.03$
  (centre-right), ${\cal M}_{\rm r}=0.1$ (right).  Each case was
  performed with different numerical resolutions to test for
  convergence: $250\times 45$ and $q_{\rm visc}=4$ (long-dashed),
  $500\times 91$ and $q_{\rm visc}=4$ (triple-dot-dashed), $1000\times
  183$ and $q_{\rm visc}=4$ (dot-dashed), $1156\times 365$ and $q_{\rm
    visc}=4$ (solid).  The horizontal dotted line indicates 98 percent
  of the flux predicted by the Goldreich \& Tremaine formula (equation
  \ref{gt}).  The value of 98 percent is that predicted by linear
  theory to be deposited in the \fe mode.  As can be seen from the
  figure, the hydrodynamic results are in excellent agreement with
  linear theory for the low-amplitude cases.  In the high-amplitude
  cases, the radial energy flux is attenuated by dissipation.}
\end{figure*}

We turn now to the non-linear hydrodynamic calculations performed with
ZEUS-2D.  Figure \ref{PF_ep0.10} plots the magnitude of the vertically
integrated radial energy flux, \fr (equation \ref{fluxeq}) as a 
function of $r$.  We consider 
four values of the Mach number near resonance ${\cal M}_{\rm r}$.
The calculations with ${\cal M}_{\rm r}=0.001$ do not have sufficient
resolution to follow the \fe mode ($r>1$) until physical dissipation
occurs.  This can be seen by the absence of convergence of the flux in
Figure \ref{PF_ep0.10}.  Because of the higher concentration of wave
energy compared with the vertically isothermal case, it is more
difficult to provide adequate resolution.  However, it is clear that
the wave propagates to at least $r=2.3$ with little dissipation, over a
distance that is many times greater than the disc thickness $H$.
Notice that the radial energy flux is negative for $r<0.7$.  This
effect is due the inwardly propagating \re mode that is excited along
with the \fe mode.  As predicted by linear theory, most of the flux
goes into an outwardly propagating \fe mode.  The magnitude of this
flux is indistinguishable from the linear prediction.  We find
$\approx 1.2$\% goes into an inwardly propagating \re mode (Figure
\ref{PF_ep0.10}, ${\cal M}_{\rm r}=0.001$), similar to the linear
prediction.

The vertical thermal stratification of the polytropic disc has a
strong effect on the distribution of wave energy of the \fe and 
\re modes compared to the
isothermal disc (Figure \ref{PF_ep0.10_cont}, top left).  As predicted
by linear theory, instead of occupying the entire thickness of the
disc, as in the isothermal case, the energy in the \fe mode is
channelled towards the surface of the disc.  It effectively becomes a
wave that travels along the surface of the disc.  For the \re mode,
rather than being channelled towards the mid-plane, it continues to
occupy the entire thickness of the disc as it propagates towards the
central object. This behaviour differs from the vertically isothermal
case, where the \re mode is channelled towards the mid-plane.  The
reason for this difference is that the channelling of the \re mode is
due to vertical buoyancy (Lubow \& Pringle 1993). The difference in
behaviours is due to the vanishing buoyancy frequency throughout the
vertically polytropic disc, unlike the vertically isothermal case.

The wave energy and dissipation for our weakest driving calculation,
${\cal M}_{\rm r}=0.001$, are shown in Figure \ref{PF_ep0.10_cont}
(left panels). The dissipation is only suggestive, since numerical
convergence has not occurred at larger radii.  However, the 
channelling of the \fe mode to the surface of the disc is obvious.  
The wave dissipates at $r\approx 2.3$ in the highest resolution
calculation ($N_r\times N_{\theta}=1156\times 365$) and at $r\approx 2.2$
with half of this resolution ($N_r\times N_{\theta}=1000\times 183$).

The low-amplitude \re mode propagates towards the centre of the disc
(Figure \ref{PF_ep0.10_cont} (left
panels).  The zig-zag pattern that can be seen in the distribution of
wave energy is due to the superposition of the \re mode (for which the
wave energy has a minimum on the mid-plane) and the evanescent tail of
the \fe mode.  The wavelength of the \re mode is equal to the
wavelength of this zig-zag pattern and decreases rapidly with
decreasing radius.  Linear theory predicts that the wavelength should
be $\propto r^{5/2}$ and this is confirmed by the non-linear
calculations.  As the wave propagates towards the centre of the disc
it dissipates numerically because the wave becomes unresolved (the
size of the grid zones decreases logarithmically, while the wavelength
decreases more rapidly).

\begin{figure*}
\vspace{12.24truecm}
\caption{\label{PF_ep0.10_cont}  The case considered here is the \fe mode
  in a polytropic disc with $H/r=0.1$ and $\tau=25000$.  The kinetic
  energy density in the wave ($\log E$, equation \ref{energyeq}) is
  shown in the upper panels.  The energy dissipation rate per unit
  volume is shown in the lower panels as a logarithmic grey-scale.
  The grey-scales cover four orders of magnitude.
  The left panels are for amplitude ${\cal M}_{\rm r}=0.001$ and the
  right panels are for amplitude ${\cal M}_{\rm r}=0.1$.  At low
  amplitude the \fe mode propagates outwards from $r=1$ and becomes
  increasingly channelled to $z\approx H$.  The dissipation occurs at
  $r\approx 2.2$ as the wave becomes non-linear (upper panel, Figure
  \ref{PF_ep0.10_peak_b}).  In the high-amplitude case, wave
  channelling of the \fe mode still occurs but the high amplitude
  results in dissipation closer to the resonance.  In both cases, the
  pattern visible at $r<1$ is generated by a superposition of the
  evanescent tail of the \fe mode and a low-amplitude inwardly
  propagating r~mode.
  In both cases the grid resolution is $N_r\times N_{\theta}=1000 \times 183$
  and $q_{\rm visc}=4.0$.  }
\end{figure*}

\subsubsection{Dependence on wave amplitude}

As described above, the resolution of the calculations with ${\cal
  M}_{\rm r}=0.001$ is not sufficient to resolve physical dissipation
of the \fe mode due to shocks.  However, calculations with ${\cal
  M}_{\rm r} \ge 0.01$ are able to resolve the \fe mode until physical
dissipation occurs.  Considering the convergence of \fr in Figure
\ref{PF_ep0.10}, we find that physical dissipation is marginally
resolved in the ${\cal M}_{\rm r}=0.01$ calculations with $N_r\times
N_{\theta}=1156\times 365$ (there is very little difference in the
radius at which \fr rapidly decreases between this calculation and
that with $N_r\times N_{\theta}=1000\times 183$).  Calculations with
${\cal M}_{\rm r} \ge 0.03$ are easily resolved with $N_r\times
N_{\theta}=1000\times 183$.  The radii to which these waves 
propagate before physical dissipation occurs are given in 
Table \ref{table1}.

\begin{table}
\centerline{
\begin{tabular}{l|cccc}
\hline
$H/r$ 			& 0.05	& 0.1 	& 0.2	\\
\hline
${\cal M}_{\rm r}$=0.01	& 1.7	& 1.8 	& 1.9+ 	\\
${\cal M}_{\rm r}$=0.03	& 1.5	& 1.4	& 1.5 	\\
${\cal M}_{\rm r}$=0.1	& 1.4	& 1.4	& 1.4	\\
\hline
\end{tabular}}
\caption{\label{table1} The radius $r$ at which the \fe mode dissipates
is tabulated for polytropic discs in which the calculations
resolve the shock dissipation.  Values are given for different disc
thicknesses, $H/r$, and for different initial wave amplitudes
${\cal M}_{\rm r}$.  The `+' indicates that this value is a lower limit.
Note that the radius of dissipation is almost independent of the disc
thickness as predicted by the theory of wave channelling.}
\end{table}

Compared with the case of the vertically isothermal disc, the wave
propagates a shorter distance before dissipating, although still much
greater than $H$.  The increased non-linearity is due to the increase
in the magnitude of velocity perturbations and the decrease in local
sound speed that occurs as the wave is channelled into the small
region near the disc surface.

\begin{figure}
\centerline{\psfig{figure=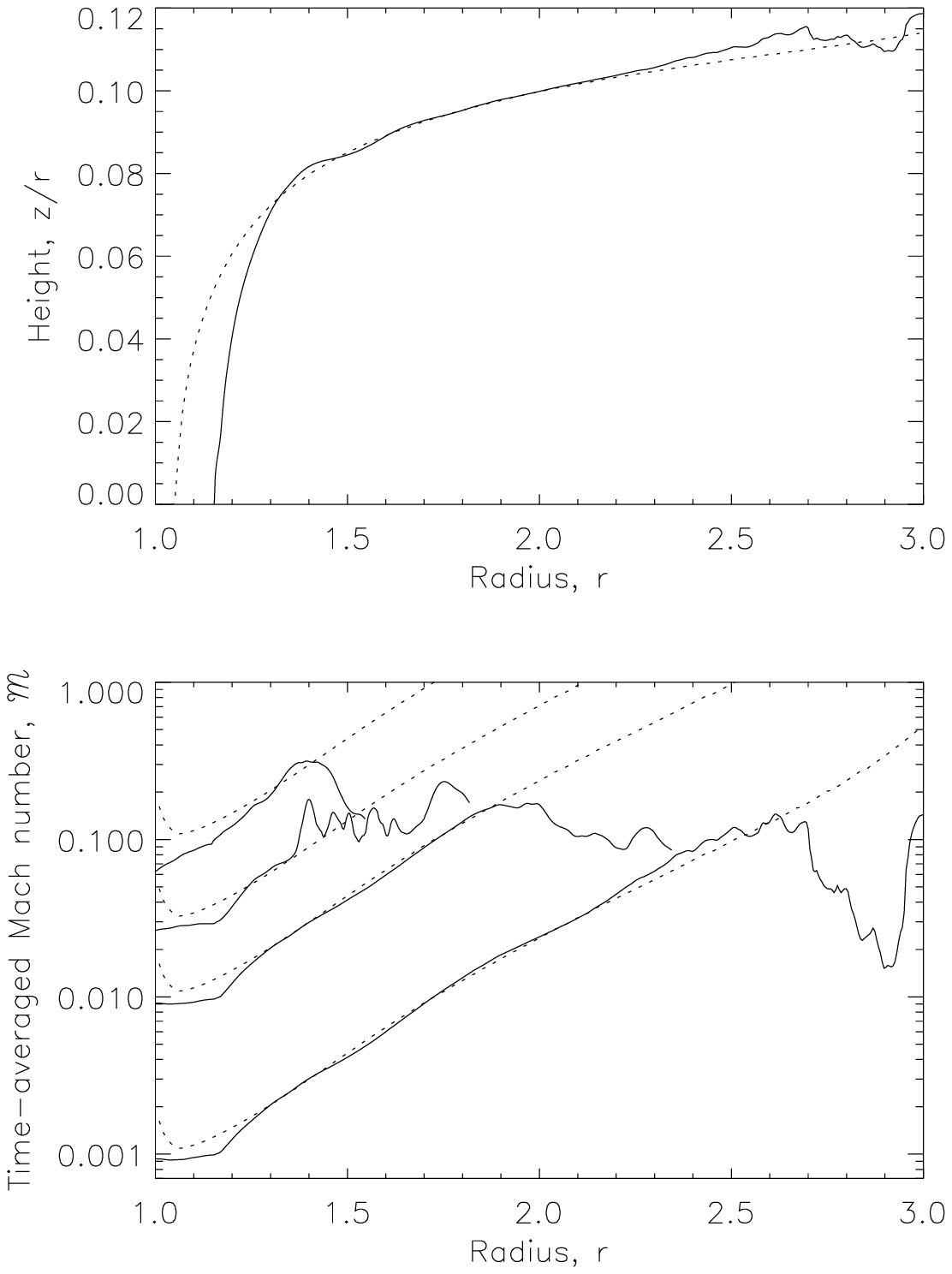,width=8.0truecm,height=10.66truecm,rwidth=8.0truecm,rheight=10.66truecm}}
\caption{\label{PF_ep0.10_peak_a} Comparison of the non-linear f$^{\rm e}$ 
  mode with linear theory for the polytropic disc with $H/r=0.1$ and
  $\tau=25000$.  Top panel: We plot as a function of radius, the
  height, $z$, in the disc where the wave energy density reaches a
  maximum.  The linear prediction is given by the dotted line and the
  low-amplitude ${\cal M}_{\rm r}=0.001$ computation is given by the
  solid line.  For higher amplitude waves the maxima
  lie roughly along the same curve, but attempting to extract them
  from the calculations results in curves that are so noisy that plotting
  them makes the figure unintelligible.  
  Bottom panel: The time-averaged wave Mach number at the
  location of the peak in wave energy density (as defined in the upper
  panel) is plotted as a function of radius.  Four cases are shown:
  non-linear calculations with (lower to upper solid lines) 
  ${\cal M}_{\rm r}=0.001$, ${\cal M}_{\rm r}=0.01$, ${\cal M}_{\rm r}=
  0.03$, ${\cal M}_{\rm r}=0.1$ and the corresponding linear
  predictions.  The numerical results display excellent agreement with
  linear theory both for the location of the maximum wave energy and
  for the Mach number of the wave at these maxima, until the wave
  dissipates in shocks (${\cal M}_{\rm r}=0.01$, ${\cal M}_{\rm r}=0.03$, 
  ${\cal M}_{\rm r}=0.1$) or numerical dissipation 
  (${\cal M}_{\rm r}=0.001$). }
\end{figure}

\begin{figure}
\centerline{\psfig{figure=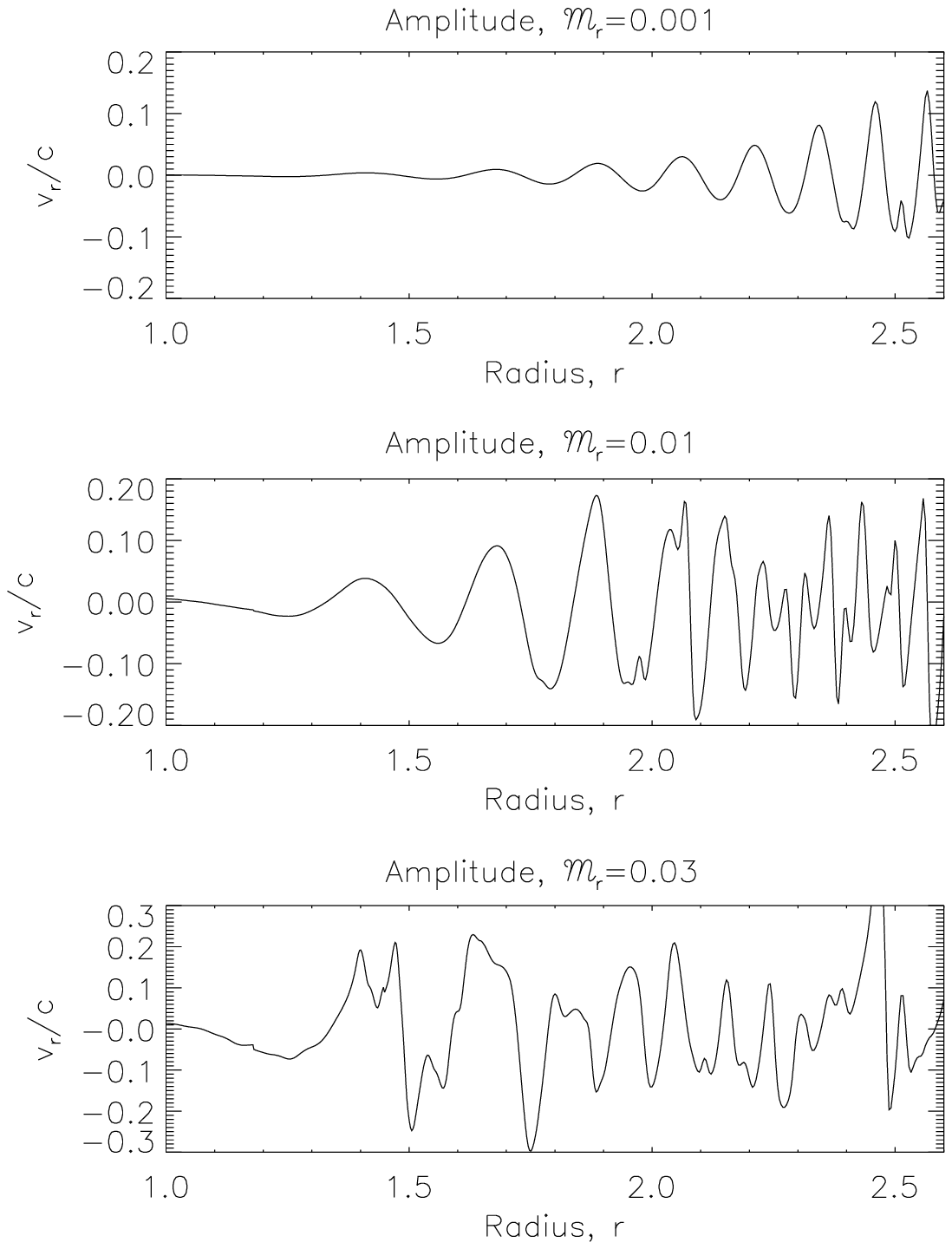,width=8.0truecm,height=10.0truecm,rwidth=8.0truecm,rheight=10.0truecm}}
\caption{\label{PF_ep0.10_peak_b} The Mach number of the radial motion
  of the wave at the height in the disc at which the wave energy
  density peaks (see Figure \ref{PF_ep0.10_peak_a}) is plotted at a
  particular instant as a function of radius for the f$^{\rm e}$ mode
  in the polytropic disc with $H/r=0.1$ and $\tau=25000$.  Various
  wave amplitudes are shown: ${\cal M}_{\rm r}=0.001$ (top), ${\cal
    M}_{\rm r}=0.01$ (middle) and ${\cal M}_{\rm r}=0.03$ (bottom).
  For the lowest amplitude case, ${\cal M}_{\rm r}=0.001$, the wave
  maintains its sinusoidal profile, but with increasing amplitude as
  it is channelled towards the surface, until numerical dissipation
  occurs.  For the higher amplitudes, ${\cal M}_{\rm r}=0.01$ and
  ${\cal M}_{\rm r}=0.03$ the wave becomes non-linear very rapidly,
  shocks and dissipates.}
\end{figure}

In Figure \ref{PF_ep0.10_peak_a}, we plot the height above the
mid-plane of the peak of the wave energy density as a function of
radius and compare it with the prediction from linear theory.  We also
plot the mean Mach number of the wave along this peak in wave energy
${\cal M}(r, \theta)$ and compare this to linear theory.  Away from
the resonance ($r \gsim 1.3$), where the linear theory should give an
accurate description, the agreement between the linear theory and the
non-linear hydrodynamic calculations is excellent for both the
location of the peak in the wave energy density and the Mach number of
the wave along this maximum.  It appears that shock dissipation occurs
once the mean Mach number ${\cal M}$ exceeds approximately 0.15 (i.e.
the peak-to-trough mean Mach number exceeds $\approx 0.4$).  Note that
the Mach number reached by the ${\cal M}_{\rm r}=0.001$ calculation at
$r=2.5$ is close to this value, implying that even the
lowest-amplitude wave may be marginally resolved with the resolution
of $N_r\times N_{\theta}=1156\times 365$.

Finally, we note that the efficiency of the dissipation depends on
${\cal M}_{\rm r}$, in that stronger waves result in significant flux
beyond the dissipation radius (cf. ${\cal M}_{\rm r}=0.01$ and ${\cal
  M}_{\rm r}=0.1$ in Figure \ref{PF_ep0.10}).  This might be because
the strong wave dissipation modifies the equilibrium disc structure in
a manner that makes non-linear dissipation more difficult.

\begin{figure*}
\centerline{\psfig{figure=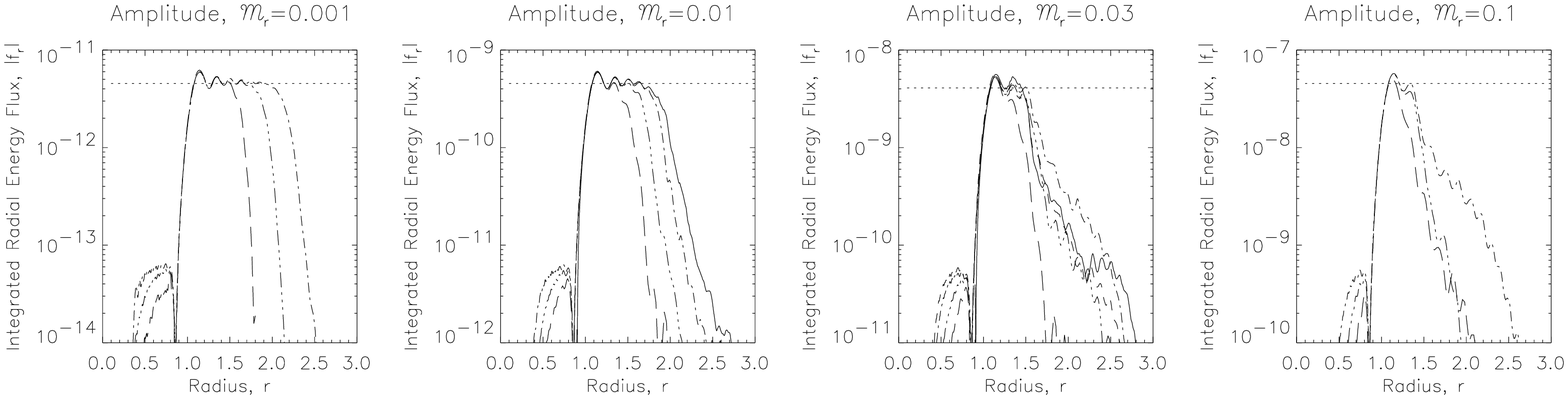,width=17.5truecm,height=4.362truecm,rwidth=17.5truecm,rheight=4.362truecm}}
\caption{\label{PF_ep0.05} The integrated radial energy flux,
  $|f_{\rm r}|$,  is plotted 
  as a function of radius for the \fe mode in a polytropic disc with
  $H/r=0.05$ and $\tau=25000$.  The results are shown for four
  different wave amplitudes: ${\cal M}_{\rm r}=0.001$ (left), ${\cal
    M}_{\rm r}=0.01$ (centre-left), ${\cal M}_{\rm r}=0.03$
  (centre-right), ${\cal M}_{\rm r}=0.1$ (right).  Each case was
  performed with different numerical resolutions to test for
  convergence: $250\times 45$ and $q_{\rm visc}=4$ (long-dashed),
  $500\times 91$ and $q_{\rm visc}=4$ (triple-dot-dashed), $1000\times
  183$ and $q_{\rm visc}=4$ (dot-dashed), $1156\times 365$ and $q_{\rm
    visc}=4$ (solid).  The horizontal dotted line indicates 98 percent
  of the flux predicted by the Goldreich \& Tremaine formula (equation
  \ref{gt}).  The value of 98 percent is that predicted by linear
  theory to be deposited in the \fe mode.  As can be seen from the
  figure, the hydrodynamic results are in excellent agreement with
  linear theory for the low-amplitude cases.  In the high-amplitude
  cases, the radial energy flux is attenuated by dissipation.}
\end{figure*}

\begin{figure*}
\centerline{\psfig{figure=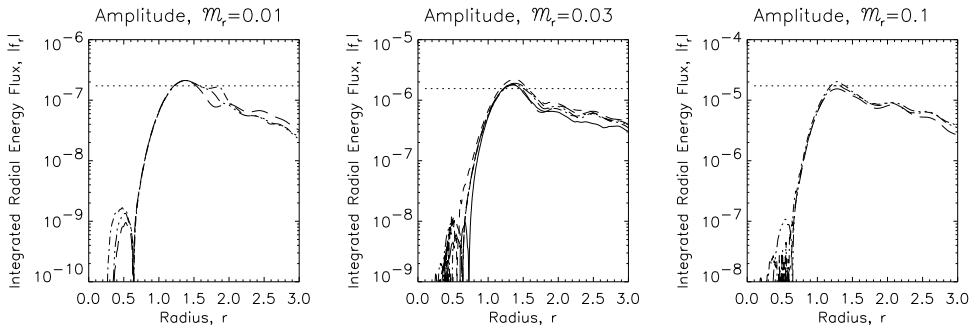,width=13.125truecm,height=4.362truecm,rwidth=13.125truecm,rheight=4.362truecm}}
\caption{\label{PF_ep0.20} The integrated radial energy flux,
  $|f_{\rm r}|$,  is plotted 
  as a function of radius for the \fe mode in a polytropic disc with
  $H/r=0.2$ and $\tau=25000$.  The results are shown for three
  different wave amplitudes: ${\cal M}_{\rm r}=0.01$ (left), ${\cal
    M}_{\rm r}=0.03$ (centre), ${\cal M}_{\rm r}=0.1$ (right).  Each
  case was performed with different numerical resolutions to test for
  convergence: $250\times 45$ and $q_{\rm visc}=4$ (long-dashed),
  $500\times 91$ and $q_{\rm visc}=4$ (triple-dot-dashed), $1000\times
  183$ and $q_{\rm visc}=4$ (dot-dashed), $1156\times 365$ and $q_{\rm
    visc}=4$ (solid).  The horizontal dotted line indicates 98 percent
  of the flux predicted by the Goldreich \& Tremaine formula (equation
  \ref{gt}).  The value of 98 percent is that predicted by linear
  theory to be deposited in the \fe mode.  As can be seen from the
  figure, the hydrodynamic results are in reasonable agreement with
  linear theory for the most highly resolved low-amplitude case.  In
  the high-amplitude cases, the radial energy flux is attenuated by
  dissipation. }
\end{figure*}

\subsubsection{Dependence on disc thickness}

\label{DependenceOnThickness}

We have seen that in a polytropic disc the \fe mode is channelled
towards the surface of the disc and dissipates due to wave steepening
when the mean Mach number at the location where the wave energy is
concentrated exceeds ${\cal M}\approx 0.15$.  We examine here how this
behaviour depends on the thickness of the disc. Recall also that that
the linear theory has been developed under a thin-disc approximation.

Figures \ref{PF_ep0.05} and \ref{PF_ep0.20} correspond to Figure
\ref{PF_ep0.10}, but for values of $H/r$ equal to 0.05 and 0.2
respectively.  In Table \ref{table1}, we give the radius of
dissipation of the \fe mode for those calculations in which the wave
dissipates due to shocks (rather than numerically).  Although $H/r$ is
varied by a factor of 4, the radius at which the wave dissipates is
almost independent of $H/r$.  This result is consistent with the
expectations of wave channelling, since channelling occurs over a
distance that is independent of $H/r$ (Lubow \& Ogilvie 1998).

The main difference between discs of different thickness is that
thinner discs seem to result in more efficient dissipation of the
wave.  For example, consider the calculations with 
${\cal M}_{\rm r}=0.01$.  The radial energy flux drops by 
nearly a factor of $\sim 1000$ between 
$r=1.7$ and $r=2.5$ for the $H/r=0.05$ disc, whereas it
only drops by a factor of $\approx 10$ for $H/r=0.1$ and only a
factor of $\approx 3$ for $H/r=0.2$.  The reason for this dependence
of the dissipation efficiency on the disc thickness is not clear.

In summary, linear theory, although derived for thin discs, gives a
good description of the wave propagation even in discs as hot as
$H/r=0.2$, although the dissipation of the waves may be less
efficient for thicker discs.

\subsection{p$^{\rm e}_{1}$ mode}

\subsubsection{Linear theory}

As discussed in Section \ref{excitation}, we adopt a driving force per
unit mass that is purely in the $\theta$-direction of the form
$a_{\theta}(r, \theta, t) = {\cal A}(r) r\cos\theta\sin\omega t$, in
order to efficiently excite the \pe mode.  As described earlier, the
\pe mode is launched at a vertical resonance at $r=1.39$ and
propagates outwards.  No other wave is expected and the radial 
energy flux of the \pe mode is predicted in Appendix B2.  
In a vertically polytropic disc, the \pe mode energy is expected 
to channel to the disc surface.

\begin{figure*}
\centerline{\psfig{figure=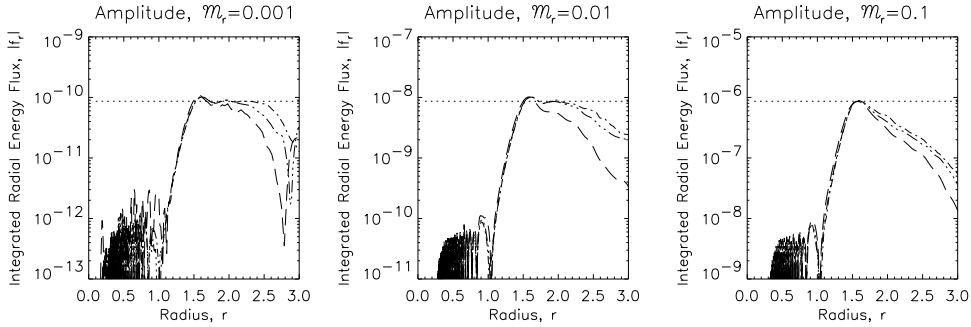,width=13.125truecm,height=4.362truecm,rwidth=13.125truecm,rheight=4.362truecm}}
\caption{\label{PP_ep0.10} The integrated radial energy flux,
  $|f_{\rm r}|$,  is plotted 
  as a function of radius for the \pe mode in a polytropic disc with
  $H/r=0.1$ and $\tau=25000$.  The results are shown for three
  different wave amplitudes: ${\cal M}_{\rm r}=0.001$ (left), ${\cal
    M}_{\rm r}=0.01$ (centre), ${\cal M}_{\rm r}=0.1$ (right).  Each
  case was performed with different numerical resolutions to test for
  convergence: $250\times 45$ and $q_{\rm visc}=4$ (long-dashed),
  $500\times 91$ and $q_{\rm visc}=4$ (triple-dot-dashed), $1000\times
  183$ and $q_{\rm visc}=4$ (dot-dashed).  The horizontal dotted line
  indicates the flux predicted by linear theory (equation
  \ref{linear_flux_p}).  As can be seen from the figure, the
  hydrodynamic results are in excellent agreement with linear theory
  for the low-amplitude cases.  At high amplitude, the radial energy
  flux is attenuated by dissipation.}
\end{figure*}

\begin{figure*}
\vspace{12.24truecm}
\caption{\label{PP_ep0.10_cont}  The case considered here is the \pe mode
  in a polytropic disc with $H/r=0.1$ and $\tau=25000$.  The kinetic
  energy density in the wave ($\log E$, equation \ref{energyeq}) is
  shown in the upper panels covering three orders of magnitude.  
  The energy dissipation rate per unit
  volume is shown in the lower panels as a logarithmic grey-scale
  covering four orders of magnitude.
  The left panels are for amplitude ${\cal M}_{\rm r}=0.001$ and the
  right panels are for amplitude ${\cal M}_{\rm r}=0.1$.  At low
  amplitude the \pe mode propagates outwards from $r\approx 1.39$ and
  becomes increasingly channelled to $z\approx H$.  The mode displays
  vertical structure as predicted by linear theory (cf.\ the
  vertically isothermal disc, Figure \ref{IP_ep0.10_cont}).  The
  dissipation occurs at $r\approx 2.5$ as the wave becomes non-linear.
  In the high-amplitude case, wave channelling of the \pe mode still
  occurs but the high amplitude results in dissipation closer to the
  resonance.  In both cases, the pattern visible at $r<1$ is generated
  by a superposition of the evanescent tail of the \fe mode and a
  low-amplitude inwardly propagating r~mode.
  In both cases the grid resolution is $N_r\times N_{\theta}=1000 \times 183$
  and $q_{\rm visc}=4.0$.}
\end{figure*}

\begin{figure*}
\centerline{\psfig{figure=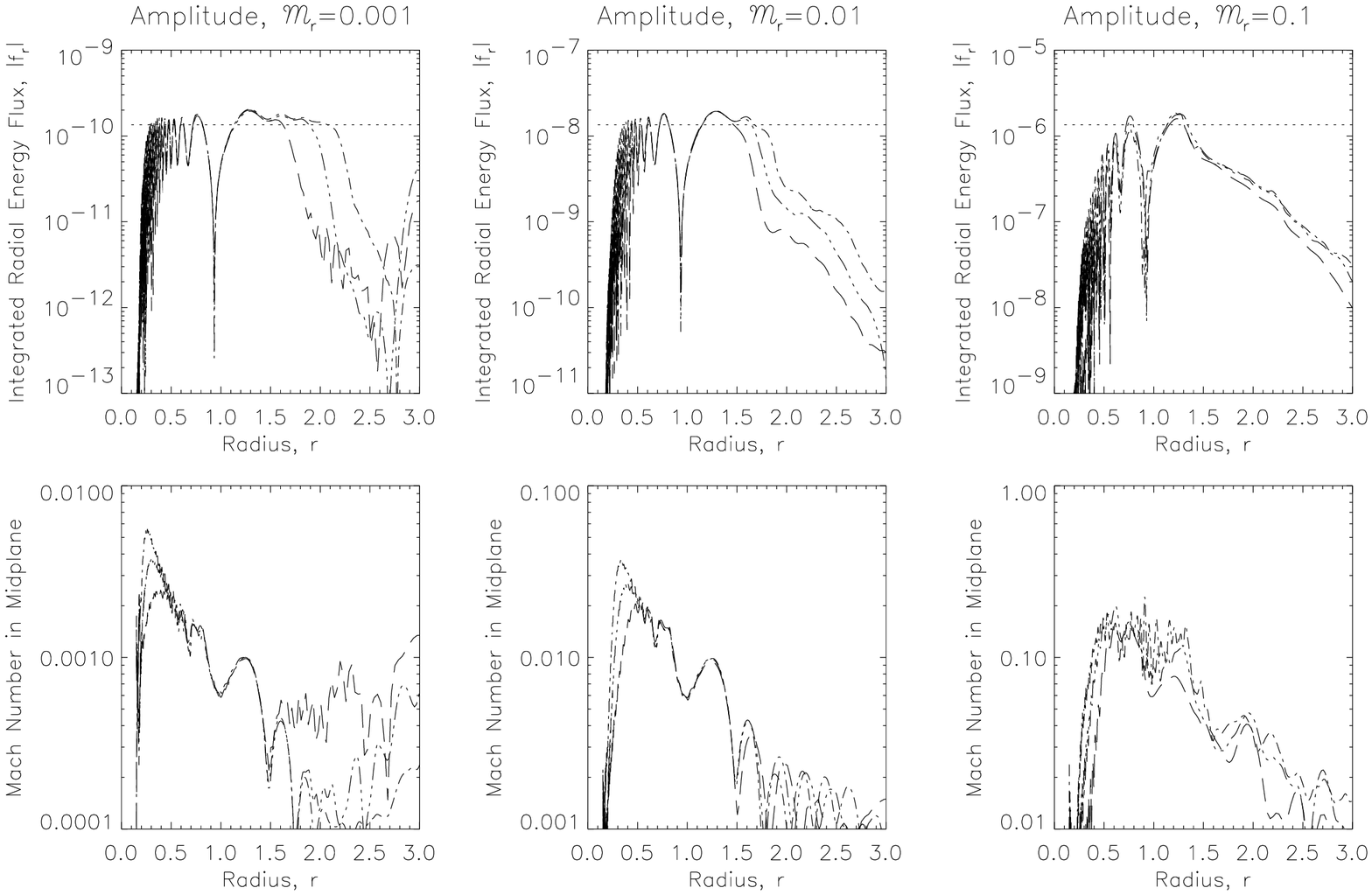,width=13.125truecm,height=8.75truecm,rwidth=13.125truecm,rheight=8.75truecm}}
\caption{\label{PT_ep0.10}  The integrated radial energy flux,
  $|f_{\rm r}|$,  (top) and 
  mean (time-averaged) Mach number at the mid-plane of the disc, 
  ${\cal M}(r, \theta=\pi/2)$,
  (bottom) are plotted as functions of radius for the \foro
  (corrugation) mode in a polytropic disc with $H/r=0.1$ and
  $\tau=25000$.  The results are shown for three different wave
  amplitudes: ${\cal M}_{\rm r}=0.001$ (left), ${\cal M}_{\rm r}=0.01$
  (centre), ${\cal M}_{\rm r}=0.1$ (right).  Each case was performed
  with different numerical resolutions to test for convergence:
  $250\times 45$ and $q_{\rm visc}=4$ (long-dashed), $500\times 91$
  and $q_{\rm visc}=4$ (triple-dot-dashed), $1000\times 183$ and
  $q_{\rm visc}=4$ (dot-dashed).  The horizontal dotted line indicates
  the flux predicted by linear theory (equation
  \ref{linear_flux_foro}).  For this case, the inwardly propagating
  \ro mode ($r<1$) and the outwardly propagating \fo mode ($r>1$)
  carry equal amounts of energy.  As can be seen from the figure, the
  hydrodynamic results are in reasonable agreement with linear theory
  for the low-amplitude high-resolution case.  At high amplitude, the
  radial energy flux is attenuated by dissipation.  It can be seen
  that the Mach number of the \ro mode increases as it propagates
  inwards caused by the rapid decrease of the group velocity (or
  radial wavelength, Figure \ref{PT_ep0.10}).}
\end{figure*}

\begin{figure*}
\vspace{12.24truecm}
\caption{\label{PT_ep0.10_cont}  The case considered here is the \foro
  (corrugation) mode in a polytropic disc with $H/r=0.1$ and
  $\tau=25000$.  The kinetic energy density in the wave ($\log E$,
  equation \ref{energyeq}) is shown in the upper panels.  The energy
  dissipation rate per unit volume is shown in the lower panels as a
  logarithmic grey-scale.  The grey-scales cover four orders of magnitude.
  The left panels are for amplitude ${\cal
    M}_{\rm r}=0.001$ and the right panels are for amplitude ${\cal
    M}_{\rm r}=0.1$.  At low amplitude the \fo mode propagates
  outwards from $r=1$ and becomes increasingly channelled to $z\approx
  H$.  Dissipation of the \fo mode occurs at $r\approx 2.2$ as the
  wave becomes non-linear.  In the high-amplitude case, wave
  channelling of the \fo mode still occurs but the high amplitude
  results in dissipation closer to the resonance.  In both cases, the
  pattern visible at $r<1$ is generated by a superposition of the
  evanescent tail of the \fo mode and the \ro mode that propagates
  inwards from $r=1$.  Note that as the \ro mode propagates inwards
  its radial wavelength (and hence its group velocity) decreases.
  In both cases the grid resolution is $N_r\times N_{\theta}=1000 \times 183$
  and $q_{\rm visc}=4.0$.}
\end{figure*}

\subsubsection{Low-amplitude, non-linear results}

Figure \ref{PP_ep0.10} plots the magnitude of the vertically
integrated radial energy flux, \fr (equation \ref{fluxeq}) 
as a function of $r$.  We consider three values of the 
mean Mach number near resonance ${\cal M}_{\rm r}$.  
The wave excitation is similar to that in the isothermal disc
with most of the energy going into an outwardly propagating \pe mode
which is excited at $r=1.39$ and a small fraction producing an
inwardly propagating \re mode which is excited at $r=1$.  We find that
the ratio of the two fluxes is $\approx 1.1$\% (Figure
\ref{PP_ep0.10}, ${\cal M}_{\rm r}=0.01$).  As noted previously, this
effect is of order $(H/r)^2$ and should vanish in the limit of a thin
disc.

In the isothermal disc, the peak of wave energy of the \pe mode in the
vicinity of the resonance ($r=1.39$) occurs off the mid-plane for $1.0
\lsim r \lsim 1.7$. (There is one maximum above, and one below the
mid-plane).  This structure changes as the wave propagates to larger
radii to a peak on the mid-plane and two off-mid-plane peaks for $r
\gsim 1.7$.  In the polytropic disc (Figure \ref{PP_ep0.10_cont}, top
left), this general structure is repeated, but with the added
complexity of the wave simultaneously being channelled to the disc
surface in a similar manner to the \fe mode.  Thus, in the vicinity of
the resonance ($1.0 \lsim r \lsim 1.7$) the wave energy has a minimum
on the mid-plane as in the isothermal disc.  At $r \gsim 1.7$, the two
off-mid-plane peaks which were present in the isothermal disc are near
the surface of the polytropic disc. The peak which was on the
mid-plane in the isothermal disc is also present in the polytropic
disc for $1.7 \lsim r \lsim 2.1$, but is rapidly channelled towards
the surface of the disc for $r \gsim 2.1$ in a manner which is similar
to the channelling of the \fe mode.  For $r \gsim 2.5$, there is very
little wave energy near the mid-plane of the disc.

The \re mode, which is excited at $r=1$ and propagates inwards,
behaves in the same way as the \re mode which was excited by the
forcing for the \fe mode.  It occupies the entire thickness of the
disc and shows no sign of physical dissipation as it propagates to the
centre.  It is dissipated numerically because of the rapid decrease of
wavelength.

\subsubsection{Wave amplitude}

We only have sufficient resolution to find the radius at which 
physical dissipation occurs for the case with ${\cal M}_{\rm r}= 0.1$ 
(Figure \ref{PP_ep0.10}).  However, in cases with the same radial
resolutions, the radius of dissipation is larger for the 
\pe mode than for the \fe mode (c.f.\ Figure \ref{PF_ep0.10})
because the wave is excited at $r=1.39$ instead of $r=1$.

\subsection{f$^{\rm o}$/r$^{\rm o}_{1}$ mode}

\subsubsection{Linear theory}

As discussed in Section \ref{excitation}, we adopt a driving force per
unit mass that is purely in the $\theta$-direction of the form
$a_\theta(r, t) = {\cal A}(r)\sin\omega t$, in order to efficiently
excite the \foro mode.  The general properties of the waves were
described in Section \ref{foromode} for the vertically isothermal
case.  In the polytropic case, the primary new feature is that wave
channelling occurs.

\subsubsection{Low-amplitude, non-linear results}

Figure \ref{PT_ep0.10} plots the magnitude of the vertically
integrated radial energy flux, \fr (equation \ref{fluxeq}), and the
mean Mach number (equation \ref{macheq}) in the mid-plane of the disc,
${\cal M}(r, \theta=\pi/2)$, as functions of $r$.  They are plotted
for three values of the mean Mach number near resonance 
${\cal M}_{\rm r}$.  The wave excitation is 
similar to that in the isothermal disc
with 50\% of the flux going into each of the inwardly and outwardly
propagating modes.

In the isothermal disc, the wave energy of the \fo mode peaked off the
mid-plane.  In the polytropic disc (Figure \ref{PT_ep0.10_cont}, top
left), this is apparent near the resonance ($1 \lsim r \lsim 1.3$),
but for $r \gsim 1.3$, channelling of the wave to the disc surface
dominates the distribution of wave energy.  In fact, apart from the
minimum in the wave energy on the mid-plane for $1 \lsim r \lsim
1.3$, the distribution of wave energy is almost identical to that of
the \fe mode (cf. Figure \ref{PF_ep0.10_cont}, top left).

The \ro mode, which is excited at $r=1$ and propagates inwards,
occupies the entire thickness of the disc as did the \re mode which
was excited by the forcing for the \fe and \pe modes.  
The mean Mach number at mid-plane increases rapidly with
decreasing radius for the \ro mode (Figure \ref{PT_ep0.10},
lower panels).  In the lowest amplitude, highest resolution 
calculation, it increases by a factor of $\approx 4$ in going 
from the resonant radius $r=1$ to radius $r=0.3$.  In reality,
this would result in the formation of shocks and dissipation of the
wave as it approaches the centre of the disc.  However, in the
calculations presented here, we do not have sufficient resolution to
resolve physical dissipation as can be seen by the lack of convergence
of \fr with increasing resolution (Figure \ref{PT_ep0.10}).  Instead,
the \ro mode is dissipated numerically because of the rapid decrease
of wavelength.

\begin{figure*}
\vspace{15.60truecm}
\caption{\label{tau} We consider the \fe mode in polytropic discs with 
  $H/r=0.1$ and varying optical depth.  From top to bottom: $\tau=5$,
  $\tau=10$, $\tau=30$, and $\tau=100$.  In each case, the amplitude
  is ${\cal M}_{\rm r}=0.01$.  The kinetic energy density in the wave (log
  E, equation 5) is shown in the left-hand panels.  The energy
  dissipation rate per unit volume is shown in the middle panels as a
  logarithmic grey-scale.  The grey-scales cover four orders of magnitude.
  The right panels give the integrated radial
  energy flux as functions of radius.  The horizontal dotted lines
  indicate the flux predicted by linear theory for the \fe mode.  As
  can be seen from the figure, the predicted fluxes for discs of
  intermediate optical depth are in excellent agreement with linear
  theory.  As the optical depth increases, there is a gradual change
  from the vertically isothermal behaviour (i.e. the wave occupying
  the entire thickness of the disc, top) to wave channelling (bottom).
  This is predicted by linear theory (Ogilvie \& Lubow 1999).  As wave
  channelling becomes more important, the attenuation of the wave
  increases.  In each case the grid resolution is $N_r\times
  N_{\theta}=500 \times 183$ and the disc is modelled over 6 vertical
  scale-heights.}
\end{figure*}

\subsubsection{Wave amplitude}

For a given wave amplitude, it appears that the \fo mode dissipates at
about the same radius as the \fe mode.  This confirms what was found
in comparing the distance of propagation of the \fe and \pe modes,
namely that the channelling of the waves towards the disc surface and
their resulting radius of dissipation occurs at about the same
distance from resonance. This result is probably a consequence of the
fact that wave channelling occurs when $k H$ is greater than unity
(Lubow \& Ogilvie 1998). This condition is determined by the wave
dispersion relations and occurs at roughly the same distance from
resonance (namely $\sim r$) for the modes considered here.

\section{Discs with $\tau = 5-100$}

\label{sectau}

Thus far, we have considered vertically isothermal discs, with optical
depth $\tau\sim 0$, or polytropic discs with tenuous isothermal
atmospheres and $\tau\approx 25000$.  Wave propagation and dissipation
are quite different for these two types of discs.  In this section we
briefly examine cases of intermediate optical depth, as defined by equation
\ref{tau_def}. This has been studied in linear theory by Ogilvie \&
Lubow (1999).  They found that as the optical depth of the disc is
decreased, outwardly propagating waves are channelled into thicker
layers.  This results in a milder increase of the Mach number as the
wave propagates outwards.

In this section, we consider only the \fe mode.  We perform
calculations with discs that have $\tau=5, 10, 30$ and $100$.  Their
individual parameters were chosen so that the thickness of the 
polytropic layer relative to $r$ is the same for each disc 
($H/r \approx 0.1$).  Each calculation had ${\cal M}_{\rm r}=0.01$.

The results are provided in Figure \ref{tau}.  As predicted
by linear theory, there is a gradual change from the vertically 
isothermal behaviour (i.e.\ the wave occupying the entire 
thickness of the disc, top) to wave channelling (bottom) as 
the optical depth is increased.  For discs of increasing
optical depth, a greater fraction of the \fe mode's energy
is channelled to the surface of the disc where it dissipates.
Thus, the fraction of the \fe mode flux that propagates 
past $r\approx 2$ decreases (right panels).  The flux in the 
$\tau=10$ disc is only reduced by a factor of $\approx 2$, while
the \fe mode loses more than 80\% of its flux if $\tau \gsim 100$.
These results demonstrate that wave channelling is important even
in discs of quite low optical depth.

\section{Conclusions and discussion}

\subsection{Comparison with linear theory}

We have performed non-linear numerical simulations of the excitation,
propagation and dissipation of axisymmetric hydrodynamic waves in an
accretion disc.  The disc is Keplerian, vertically resolved and
inviscid apart from an artificial numerical bulk viscosity that is
required in order to resolve shocks.  The waves are excited by
applying a temporally periodic acceleration throughout the
computational domain, and are launched resonantly at radii where the
forcing frequency coincides with a natural oscillation frequency of
the disc.  We have examined Lindblad resonances, vertical
resonances and hybrid vertical/Lindblad resonances.  In each case the
waves propagate radially through the disc away from the resonance,
changing continuously in vertical structure as they do so, and
ultimately dissipate.

This mechanism of generating the waves is directly
comparable with the way that non-axisymmetric waves are excited in
discs by tidal forcing from an orbiting companion in a binary or
protoplanetary system.  A low-mass companion such as a
sub-Jovian planet orbiting a star exerts its tidal influence on the disc
mainly through launching \fe-mode waves at the Lindblad
resonances (Goldreich \& Tremaine 1980).  
Lindblad resonances are also important in systems of
larger mass ratio, such as circular orbit binary star systems of extreme
mass ratio (Lin \& Papaloizou 1979) and eccentric orbit binaries 
(Artymowicz \& Lubow 1994).  Related resonances play a role in the 
growth of the eccentric disc mode in superhump binaries  (Lubow 1991) 
and possibly binaries with brown dwarf secondaries 
(Papaloizou, Nelson \& Masset 2001).  
Vertical resonances are an important aspect of the tidal interaction 
in close binary stars (Lubow 1981; Ogilvie 2002), where the Lindblad
resonances are all excluded from the disc.  The corrugation-mode
resonances we have studied are the axisymmetric equivalent of the
bending-mode resonances observed in Saturn's rings 
(Shu, Cuzzi \& Lissauer 1983).

Wherever the waves are linear, our results are in excellent and
detailed agreement with the linear theory developed by Lubow \&
Pringle (1993), Korycansky \& Pringle (1995), Ogilvie (1998), Lubow \&
Ogilvie (1998) and Ogilvie \& Lubow (1999).  Elements of this theory
can also be found in the work of Loska (1986) and Okazaki, Kato \&
Fukue (1987).  The disc acts as a waveguide that allows a variety of
wave modes to propagate in the radial direction.  The modes can be
classified similarly to stellar oscillations, and we have discussed
all four classes (f, p, r and g) to some extent in this paper.  Each
mode has a dispersion relation that connects the wave frequency and
the radial wavenumber at every radius.  In general, waves of a given
frequency can propagate only in restricted radial intervals, bounded
by turning points where the wavenumber vanishes.  The turning points
correspond to the resonant radii where waves are launched by an
applied periodic forcing.

The linear theory is confirmed by the simulations in a number of
respects.  First, the concept of the disc as a waveguide that supports
discrete propagating modes is validated.  The waves are vertically
confined by the continuous stratification of the disc, not by
zero-density surfaces, artificial boundaries or other discontinuities.
The solutions exhibit the same `separation of variables' that also
features in the linear analysis.  That is, the waveform is an
oscillatory function of radius (and time) multiplied by a vertical
profile that changes slowly with radius.  The fact that the velocity
amplitudes of the linear solutions may formally diverge high in the
atmosphere does not invalidate the rest of the solution when the
energy density of the wave is vertically confined in the bulk of the
disc.  Secondly, the energy fluxes imparted to the different wave
modes through launching at different types of resonance are in
excellent agreement with the predictions of linear theory presented in
Appendix~B.  This analysis of resonant wave launching is closely
related to the well-known resonant torque formula of Goldreich \&
Tremaine (1979) and to the analysis of vertical resonances by Lubow
(1981).  Finally, the detailed waveforms predicted by linear theory
are confirmed in the simulations (see, e.g., Figure \ref{refraction}
and the next subsection), even when the disc is not especially thin.

\begin{figure*}
\vspace{18.0truecm}
\caption{\label{refraction}
  Comparison of the simulations with linear theory, and comparison of
  wave channelling with acoustic refraction.  In the upper grey-scale
  figure, we plot a snapshot of the radial velocity multiplied by the
  square root of the density, $v_{\rm r}\sqrt{\rho}$, from a 
  simulation involving an \fe mode in a polytropic disc that joins
  smoothly on to a low-mass isothermal atmosphere ($H/r=0.1$,
  $\tau=25000$). (The square of this quantity is the radial part of the
  instantaneous wave energy density, c.f.\ equation 5).  
  The amplitude of the wave is ${\cal M}_{\rm r}=0.001$ and
  the grid resolution is $N_r\times N_{\theta}=1000\times 183$.  In
  the lower grey-scale figure, we plot the corresponding
  semi-analytical prediction of the linear theory of wave channelling,
  ignoring the r~mode and using the WKB approximation in the radial
  direction (which results in a mild singularity at the Lindblad
  resonance).  The agreement is excellent until dissipation
  occurs.  Over the upper grey-scale, we trace the rays that would
  result from the refraction of an initially vertical wavefront at the
  resonant radius.  The refracted wavefronts, which are orthogonal to
  the rays, would rapidly become severely tilted and the wave would be
  predicted to propagate almost vertically into the isothermal
  atmosphere.  However, the simulation shows that the wavefronts in
  fact remain nearly vertical as the wave propagates horizontally
  along the base of the disc atmosphere (shown by the dashed line,
  $z=H$).  The \fe mode is generated throughout the thickness of the
  disc at $r=1$ but its energy rises from within the polytropic layer
  owing to wave channelling and becomes confined near the base of the
  atmosphere.  The wave propagates {\it without loss of flux\/} to
  $r\approx 2.2$ (Figure \ref{PF_ep0.10}, dot-dashed line, left
  panel) before most of its energy is dissipated in two regions at the
  base of the atmosphere (white contour lines at $r\approx 2.2$). 
  The dissipation in this example is partly numerical, as suggested 
  by Figure \ref{PF_ep0.10}.  If we were able to increase the resolution 
  further, the wave would propagate a greater distance (c.f.\ different 
  resolutions in the left panel of Figure \ref{PF_ep0.10}) and might agree 
  even more closely with the linear prediction. }
\end{figure*}

\subsection{Wave channelling versus refraction}

The continuous change of the vertical profile of a wave as it
propagates radially is a process we have termed `wave channelling'
(Lubow \& Ogilvie 1998).  The energy density of the wave can be
channelled either towards the surfaces of the disc or towards the
mid-plane as it propagates away from its site of launching.  The
results of this paper provide ample evidence for both types of wave
channelling.  In particular, the f and p~modes in a disc with a
vertical temperature gradient are channelled towards the surfaces
(e.g. Figures \ref{PF_ep0.10_cont}, \ref{PP_ep0.10_cont} and
\ref{PT_ep0.10_cont}), while the r~modes in a disc with a vertical
entropy gradient are channelled towards the mid-plane (e.g. Figures
\ref{IF_ep0.10_cont} and \ref{IT_ep0.10_cont}) as predicted by Lubow
\& Pringle (1993).

The wave energy of the \fe mode, which is the principal mode launched
at a Lindblad resonance, rises towards the surfaces of a thermally
stratified disc.  Close to the resonance, the \fe mode occupies
the full vertical extent of the disc and behaves
compressibly.  As the mode propagates away from the resonance,
its energy within the disc midplane region rises and becomes
confined (or channelled) to a layer at the base of
the disc atmosphere. The extent of channelling depends
on the degree of thermal stratification. In this regime,
the wave becomes a surface gravity wave and behaves
incompressibly (Ogilvie 1998; Lubow \& Ogilvie 1998).
{\it The process by which this occurs is wave
channelling and not acoustic refraction.} This is demonstrated as
follows.

If refraction were responsible (e.g.\ Lin et~al. 1990a,b), the wave 
would be directed upwards into the isothermal atmosphere after 
travelling a radial distance of order $H$.  This is not observed in the
simulations.  Plots of the wave energy for low-amplitude waves in
polytropic discs (Figures \ref{PF_ep0.10_cont}, \ref{PP_ep0.10_cont}
and \ref{PT_ep0.10_cont}) clearly show that the waves propagate along
the base of the atmosphere before dissipating in shocks.  Propagation
does not switch from horizontal to vertical, as would be expected if
simple refraction were involved.  In fact, the \fe mode is launched 
even in an incompressible fluid with a vertical energy distribution 
that is similar to the compressible case.  Consequently, the process 
of energy concentration cannot be due to acoustic refraction.

Figure \ref{refraction} shows an explicit comparison between the
results of the simulations and the predictions of acoustic refraction.
In order to show both the wavefronts and the vertical extent of the
wave, we plot a snapshot of the radial velocity, normalized by the RMS
velocity, and multiplied by the mean wave energy density, from a
simulation involving an \fe mode in a polytropic disc.  On top of the
grey-scale, we trace the rays that would result from the refraction of
an initially vertical wavefront at the resonant radius.  The refracted
wavefronts, which are orthogonal to the rays, would rapidly become
severely tilted and the wave would be predicted to propagate almost
vertically into the isothermal atmosphere.  However, the simulations
show that the wavefronts in fact remain nearly vertical as the wave
propagates horizontally along the base of the disc atmosphere.  The
wave propagates {\it without loss of flux\/} to $r\approx 2.2$ before
most of its energy is dissipated in two regions at the base of the
atmosphere (see Figure \ref{PF_ep0.10} for the radial energy flux).
Figure \ref{refraction} demonstrates that the linear theory of wave
channelling accurately predicts the propagation of the wave until
non-linear dissipation occurs.

In summary, our non-linear hydrodynamic calculations clearly show that
for axisymmetric waves the disc acts as a waveguide and that the 
behaviour of the waves is determined primarily by wave channelling 
and not refraction.  Explanations based on simple 
refraction are incorrect.  Ray tracing, if applied
correctly, does offer a valid description of modes with a short
vertical wavelength, i.e. those of large vertical mode number (see
Figure 11 of Lubow \& Pringle 1993).  However, such modes are unlikely
to be excited by tidal resonant forcing and the \fe mode, in particular, 
is vertically evanescent and has a vertical mode number of zero.
As mentioned in Section 1, non-axisymmetric waves of low 
azimuthal wavenumber $m\lsim r/H$ are almost indistinguishable 
from axisymmetric waves on the scale of a few $H$.  
We therefore expect that such waves will also propagate as if 
through a waveguide and be subject to similar wave channelling 
in agreement with linear theory.  Several papers (e.g.\ Lin et 
al.\ 1990a,b; Terquem 2001) have interpreted the behaviour
of low-$m$ (e.g.\ $m=2$) non-axisymmetric waves as being due 
to refraction.  Given our results for $m=0$ waves, we conclude 
that the dominant behaviour of low-$m$ non-axisymmetric waves is 
almost certainly determined by wave channelling and not refraction.

\subsection{Dissipation of waves}

In situations where waves in discs are excited by tidal forcing from
an orbiting companion, the site at which a wave ultimately dissipates
is of some importance, because the energy and angular momentum carried
by the wave are transferred to the disc there.  The simulations
provide valuable information on the location and means of wave
dissipation, matters about which we were previously able only to
speculate (e.g. Lubow \& Ogilvie 1998).

In a nearly inviscid disc, dissipation can occur only if the wave
develops very large velocity gradients.  One way to achieve this is
the classical non-linear steepening of an acoustic wave in a gas with
$\gamma>1$ (e.g. Lighthill 1978).  The crest of the wave travels
faster than the trough and the wave steepens as it propagates until
shocks form.  This is the principal effect leading to the dissipation
of the \fe mode in a vertically isothermal disc with $\gamma=5/3$
(Figure \ref{IF_ep0.10_peak}).  The situation is not entirely
straightforward because the wave Mach number is not uniform across the
wavefront.  In addition, the steepening may be accelerated by the
gradual increase of wave amplitude as the wave propagates outwards
into material of lower surface density while conserving its energy
flux.  On the other hand, the steepening may be temporarily postponed
by the dispersive character of the wave close to the resonance.

The r~modes in the vertically isothermal disc are channelled into an
increasingly thin layer near the mid-plane and simultaneously develop
very short radial wavelengths.  These modes are therefore susceptible
to viscous damping even in a nearly inviscid disc.  In our
simulations, these modes develop scales shorter than the grid spacing
and are ultimately lost.

In the polytropic disc, classical wave steepening does not occur in
any simple sense because none of the waves behaves like a plane
acoustic wave.  The \fe mode, in particular, is highly dispersive,
which tends to resist non-linear steepening.  It undergoes rapid wave
channelling to the base of the isothermal atmosphere and behaves like
a surface gravity wave.  The wave channelling enhances the amplitude
of the wave by concentrating its energy into a smaller region of
space.  In earlier work (Lubow \& Ogilvie 1998; Ogilvie \& Lubow 1999) 
we showed that this effect is typically sufficient to amplify the wave
to sonic velocities where steepening and dissipation are unavoidable.
At the same time, the radial and vertical length-scales of the wave
also decrease as it propagates outwards (Figure \ref{refraction}),
making linear viscous damping a possible source of dissipation.  
Because wave channelling occurs over a distance that is almost
independent of the thickness of the disc, the dissipation of waves 
with the same initial Mach number near the reasonance occurs at 
the same radius regardless of the disc's thickness.  We have performed 
calculations with discs whose thicknesses vary by a factor of 4 to 
demonstrate this effect (Section \ref{DependenceOnThickness} 
and Table \ref{table1}).  
The results of the simulations suggest that the \fe mode in a polytropic
disc dissipates energy at the base of the isothermal layer at a Mach
number of approximately $0.2$ (Figure \ref{PF_ep0.10_peak_a}), which
it acquires through the effect of wave channelling.  The wave energy
does not propagate vertically within the isothermal atmosphere, contrary
to the suggestion of Papaloizou \& Lin (1995).

\subsection{Outlook}

In future we hope to address some questions that remain unanswered in
this paper.  In particular, a more detailed study of the competition
between linear dispersion and non-linear steepening of waves in discs
would be valuable.  Some important wave phenomena will require
non-axisymmetric simulations.  For example, a Keplerian disc supports
slowly varying $m=1$ density and bending waves that effect an
eccentric distortion and a warping of the disc, respectively.  Unlike
those studied in this paper, these waves can propagate over long
distances without experiencing wave channelling.  Finally, attention
should be given to studying wave propagation in possibly more
realistic models such as fully turbulent discs and layered discs
(Gammie 1996).  These challenges provide ample opportunities for
further developments.

\section*{Acknowledgments}

We are grateful to Jim Stone for providing the ZEUS-2D code and for
many useful discussions.  We also gratefully acknowledge support from
NASA grants NAG5-4310 and NAG5-10732, the STScI visitor program, and
the Institute of Astronomy visitor programme.  Some of the
calculations discussed in this paper were performed on the GRAND
computer.  GIO acknowledges the support of the Royal Society through a
University Research Fellowship.

\appendix

\section{Disc models}

\subsection{A polytropic disc with an isothermal atmosphere}

We consider a differentially rotating fluid with density $\rho(R,z)$,
pressure $p(R,z)$ and angular velocity $\Omega(R,z)$, referred to
cylindrical polar coordinates $(R,\phi,z)$.  The equations of
equilibrium are
\begin{eqnarray}
  -R\Omega^2&=&-{{1}\over{\rho}}{{\partial p}\over{\partial R}}
  -{{\partial\Phi}\over{\partial R}},
  \label{horizontal}\\
  0&=&-{{1}\over{\rho}}{{\partial p}\over{\partial z}}
  -{{\partial\Phi}\over{\partial z}},
  \label{vertical}
\end{eqnarray}
where the gravitational potential is
\begin{equation}
\Phi(R,z)=-GM(R^2+z^2)^{-1/2}.
\end{equation}

Let the pressure and density be related by (cf.~Lin, Papaloizou \&
Savonije 1990)
\begin{equation}
  p=c^2\rho\left[\left({{1}\over{n+1}}\right)
  \left({{\rho}\over{\rho_{\rm s}}}\right)^{1/n}+1\right],
  \label{p}
\end{equation}
where $n$ is a positive constant, while $c(R)$ and $\rho_{\rm s}(R)$
are positive functions.  At high densities $\rho\gg\rho_{\rm s}$, the
gas becomes vertically polytropic, with polytropic index $n$.  At low
densities $\rho\ll\rho_{\rm s}$, the gas becomes vertically
isothermal, with sound speed $c$.  The transitional density $\rho_{\rm
  s}$ defines the approximate surface of the disc.

Define the pseudo-enthalpy
\begin{equation}
  h=c^2\left[\left({{\rho}\over{\rho_{\rm s}}}\right)^{1/n}+
  \ln\left({{\rho}\over{\rho_{\rm s}}}\right)-1\right].
\end{equation}
Then it follows that
\begin{equation}
  {{{\rm d}p}\over{\rho}}={\rm d}h+F\,{\rm d}R,
\end{equation}
where
\begin{eqnarray}
  F & = & {{{\rm d}(c^2)}\over{{\rm d}R}}
  \left[-\left({{n}\over{n+1}}\right)
  \left({{\rho}\over{\rho_{\rm s}}}\right)^{1/n}-
  \ln\left({{\rho}\over{\rho_{\rm s}}}\right)+2\right] \nonumber \\
  & & \mbox{} + c^2~{{{\rm d}\ln\rho_{\rm s}}\over{{\rm d}R}}
  \left[\left({{1}\over{n+1}}\right)
  \left({{\rho}\over{\rho_{\rm s}}}\right)^{1/n}+1\right].
\end{eqnarray}

Consider first the vertical equilibrium at each $R$ separately.
Equation (\ref{vertical}) becomes
\begin{equation}
  0=-{{\partial h}\over{\partial z}}-{{\partial\Phi}\over{\partial z}},
\end{equation}
with solution
\begin{equation}
  h=-\Phi+f(R).
  \label{solution}
\end{equation}
By considering the mid-plane $z=0$ we determine
\begin{equation}
  f=c^2\left[\left({{\rho_{\rm c}}\over{\rho_{\rm s}}}\right)^{1/n}+
  \ln\left({{\rho_{\rm c}}\over{\rho_{\rm s}}}\right)-1\right]
  -{{GM}\over{R}},
\end{equation}
where $\rho_{\rm c}(R)$ is the mid-plane density.  In general, therefore,
\begin{eqnarray}
\lefteqn{c^2\left[\left({{\rho}\over{\rho_{\rm s}}}\right)^{1/n}-
  \left({{\rho_{\rm c}}\over{\rho_{\rm s}}}\right)^{1/n}+
  \ln\left({{\rho}\over{\rho_{\rm c}}}\right)\right]=}\nonumber\\
  &&\qquad{{GM}\over{\sqrt{R^2+z^2}}}-{{GM}\over{R}}.
\end{eqnarray}
Given the functions $\rho_{\rm c}(R)$, $\rho_{\rm s}(R)$ and $c(R)$,
this equation can be solved numerically to determine $\rho$ at any given
point, and $p$ follows from equation (\ref{p}).

Now consider the radial equilibrium.  Equation (\ref{horizontal})
becomes
\begin{equation}
  R\Omega^2={{\partial h}\over{\partial R}}+F+
  {{\partial\Phi}\over{\partial R}}.
\end{equation}
Thus, from equation (\ref{solution}),
\begin{eqnarray}
  R\Omega^2&=&{{{\rm d}f}\over{{\rm d}R}}+F              \nonumber\\
  &=&{{GM}\over{R^2}}
  +{{{\rm d}(c^2)}\over{{\rm d}R}}
  \left[\left({{\rho_{\rm c}}\over{\rho_{\rm s}}}\right)^{1/n}-
  \left({{n}\over{n+1}}\right)
  \left({{\rho}\over{\rho_{\rm s}}}\right)^{1/n}\right]   \nonumber\\
  &&+~{{{\rm d}(c^2)}\over{{\rm d}R}}
  \left[1- \ln\left({{\rho}\over{\rho_{\rm c}}}\right)\right]
\nonumber\\
  &&+~c^2{{{\rm d}\ln\rho_{\rm s}}\over{{\rm d}R}}
  \left[\left({{1}\over{n+1}}\right)
  \left({{\rho}\over{\rho_{\rm s}}}\right)^{1/n}-
  {{1}\over{n}}
  \left({{\rho_{\rm c}}\over{\rho_{\rm s}}}\right)^{1/n}\right]
\nonumber\\
  &&+~c^2{{{\rm d}\ln\rho_{\rm c}}\over{{\rm d}R}}
  \left[{{1}\over{n}}\left({{\rho_{\rm c}}\over{\rho_{\rm
s}}}\right)^{1/n}
  +1\right],
\end{eqnarray}
which determines the angular velocity at every point.

Let us assume that the disc is thin and that $\rho_{\rm c}\gg\rho_{\rm
  s}$ so that there is a well-defined polytropic layer with an
isothermal atmosphere outside.  The scale-height of the isothermal
layer is
\begin{equation}
  H_{\rm i}\approx{{c}\over{\Omega}},
\end{equation}
while the thickness of the polytropic layer is
\begin{equation}
  H_{\rm p}\approx2^{1/2}
  \left({{\rho_{\rm c}}\over{\rho_{\rm s}}}\right)^{1/2n}H_{\rm i}.
\end{equation}

If we take $c\propto R^{-1/2}$ and $\rho_{\rm c}/\rho_{\rm s}={\rm
  constant}$ then $H_{\rm i}\propto H_{\rm p}\propto R$ so that the
scale-height of the disc increases linearly with radius.  We define
$\beta=\rho_{\rm s}/\rho_{\rm c}$ and $\epsilon$ to be the nominal
value of $H_{\rm p}/R$.  Thus
\begin{equation}
  c=2^{-1/2}\beta^{1/2n}\epsilon\left({{GM}\over{R}}\right)^{1/2}
\end{equation}
and
\begin{equation}
  \rho_{\rm s}=\beta\rho_{\rm c}.
\end{equation}
Finally, we take the profile of $\rho_{\rm c}(R)$ to be a power law
but tapered to give inner and outer edges,
\begin{equation}\label{rhoc}
  \rho_{\rm c}=R^{-\sigma}\left[\delta
  +{{1}\over{2}}\tanh\left({{R-R_1}\over{w_1}}\right)
  +{{1}\over{2}}\tanh\left({{R_2-R}\over{w_2}}\right)\right],
\end{equation}
where $\delta\ll1$ is the factor by which the density is reduced
outside the interval $R_1\la R\la R_2$ occupied by the disc, and $w_1$
and $w_2$ are the widths of the inner and outer tapers.  The
parameters of the model are then
$n,\epsilon,\beta,\sigma,\delta,R_1,R_2,w_1,w_2$.

For our standard polytropic discs, we choose $G=1$, $M=1$,
$n=1.5$, $\beta=0.01$, $\sigma=1.5$, $\delta=0.01$, $R_2=3.0$,
$w_1=R_1\epsilon$, and $w_2=R_2\epsilon$.  The value of $R_1=0.2$ for
all but the highest resolution calculations, for which $R_1=0.6$.
Results are given for three different values of the disc thickness,
$\epsilon=H_{\rm p}/R=0.05, 0.1, 0.2$.

\subsection{A vertically isothermal disc}

In the case of a vertically isothermal disc, we take
\begin{equation}
  p=c^2\rho,
\end{equation}
\begin{equation}
  h=c^2\ln\left({{\rho}\over{\rho_{\rm c}}}\right),
\end{equation}
with $c=c(R)$, $\rho=\rho_{\rm c}(R)$.  Then the analysis proceeds as
before, but now with
\begin{equation}
  F={{{\rm d}(c^2)}\over{{\rm d}R}}
  \left[-\ln\left({{\rho}\over{\rho_{\rm c}}}\right)+1\right]+
  c^2{{{\rm d}\ln\rho_{\rm c}}\over{{\rm d}R}}
\end{equation}
and
\begin{equation}
  f=-{{GM}\over{R}}.
\end{equation}
The density is obtained from
\begin{equation}
  c^2\ln\left({{\rho}\over{\rho_{\rm c}}}\right)=
  {{GM}\over{\sqrt{R^2+z^2}}}-{{GM}\over{R}},
\end{equation}
i.e.
\begin{equation}
  \rho=\rho_{\rm c}\exp\left[{{1}\over{c^2}}
  \left({{GM}\over{\sqrt{R^2+z^2}}}-{{GM}\over{R}}\right)\right].
\end{equation}
The angular velocity is given by
\begin{equation}
  R\Omega^2={{GM}\over{R^2}}+{{{\rm d}(c^2)}\over{{\rm d}R}}
  \left[-\ln\left({{\rho}\over{\rho_{\rm c}}}\right)+1\right]+
  c^2{{{\rm d}\ln\rho_{\rm c}}\over{{\rm d}R}}.
\end{equation}
To obtain a scale-height that increases linearly with radius we take
\begin{equation}
  c=\epsilon\left({{GM}\over{R}}\right)^{1/2},
\end{equation}
where, as before, $\epsilon$ is a small constant, so that the isothermal
scale-height $H=c/\Omega$ satisfies $H/R\approx\epsilon$.  As in the
polytropic case, the profile of $\rho_{\rm c}(R)$ is given by
equation \ref{rhoc}.

We use the same values of $G$, $M$, $\sigma$, $\delta$ for the
vertically isothermal discs as we use for the polytropic discs, but
take $R_1=0.5$, $R_2=10.0$, $w_1=R_1\epsilon$ and $w_2=R_2\epsilon$,
while $n$ and $\beta$ are not used.

\section{Resonant launching of axisymmetric waves in linear theory}

\subsection{The ${\rm f}^{\rm e}$ mode}

The launching of non-axisymmetric waves at a Lindblad resonance in a
three-dimensional disc has been analysed by Lubow \& Ogilvie (1998).
The total torque exerted at the resonance is identical to that
determined by Goldreich \& Tremaine (1979) using a two-dimensional
model that neglects the vertical structure of the disc.  However, in a
three-dimensional disc the torque is partitioned into different wave
modes.  In an inviscid disc, the total torque is equal to the sum of
the radial fluxes of angular momentum of the launched waves, averaged
over time, integrated azimuthally and vertically, and reckoned in the
direction of propagation at some distance from the resonance.

In the case of axisymmetric forcing no torque is exerted.  However,
the work done at the resonance can still be determined from the
formula of Goldreich \& Tremaine (1979) in an appropriate limit, using
the fact that the energy flux of each wave is equal to the angular
momentum flux multiplied by the angular pattern frequency.  For a
Keplerian disc the rate at which work is done is equal to
\begin{equation}
  {\cal F}={{\pi^2R^2{\cal A}^2\Sigma}\over{3\omega}},
  \label{gt}
\end{equation}
where $\Sigma=\int\rho\,{\rm d}z$ is the surface density, and the
applied radial acceleration is $a_R={\cal A}(R)\exp(-{\rm i}\omega
t)$.  All quantities are evaluated at the location of the resonance.
The Lindblad resonance condition $\omega=\kappa$ corresponds to
$\omega=\Omega$ in a Keplerian disc.  In an inviscid disc, ${\cal F}$
is equal to the total energy flux in the launched waves, i.e. the sum
of their radial energy fluxes, averaged over time, integrated
azimuthally and vertically, and reckoned in the direction of
propagation at some distance from the resonance.

The fraction of the energy flux imparted to any wave mode is given by
equation (45) of Lubow \& Ogilvie (1998),
\begin{equation}
  f=\Bigg|\int\rho u_R'\,{\rm d}z\Bigg|^2\Bigg/\Sigma\int\rho|u_R'|^2
  \,{\rm d}z,
\end{equation}
where $u_R'(z)$ is the eigenfunction of the radial velocity
perturbation for the mode considered, and the integrals are over the
full vertical extent of the disc.  Again, all these quantities are
evaluated at the resonance.  The imparted fraction therefore depends
on an overlap integral between the applied forcing and the modes of
the disc.

Lubow \& Ogilvie (1998) found that, in a polytropic disc, more than
$95\%$ of the flux is taken up by the ${\rm f}^{\rm e}$ mode.  The
remainder goes mainly into the ${\rm r}_1^{\rm e}$ mode.  We have
repeated the calculation for the standard polytropic model adopted in
the present paper, including the effects of the isothermal atmosphere
and the vertical boundary conditions imposed in the simulations.  We
find that $98.3\%$ and $1.6\%$ should go into the ${\rm f}^{\rm e}$
and ${\rm r}_1^{\rm e}$ modes, respectively.  We recall that the
r~modes propagate away from the resonance in the opposite direction to
the ${\rm f}^{\rm e}$ mode.  The vertical boundaries are sufficiently
distant that they have no effect on the modes or on the flux
fractions.

In a vertically isothermal disc the eigenfunctions are known
analytically from the work of Lubow \& Pringle (1993).  The radial
velocity perturbation for the two-dimensional (or \fe)
mode is of the form\footnote{There is a sign
error in the argument of the exponential in the first and second lines
of equation (37) of Lubow \& Pringle (1993).}
\begin{equation}
  u_R'\propto\exp\left[
  \left({{\gamma-1}\over{\gamma}}\right){{z^2}\over{2H^2}}\right],
\end{equation}
while, of course, $\rho\propto\exp(-z^2/2H^2)$.  We then find for the
fraction imparted to the two-dimensional (or \fe) mode
\begin{equation}
  f_{\rm 2D}=\left[\gamma(2-\gamma)\right]^{1/2}.
\end{equation}
Only in the case of isothermal perturbations ($\gamma=1$) does the
forcing match perfectly to the two-dimensional mode.  When
$\gamma=5/3$, as in the present paper, $f_{\rm
  2D}=\sqrt{5}/3\approx74.5\%$.  A further $6.0\%$ goes into the ${\rm
  r}_1^{\rm e}$ mode, and yet smaller fractions into higher-order
r~modes.  Interestingly, $18.9\%$ of the expected flux is not
accounted for by previously documented modes, which evidently {\it do
  not form a complete set}.

When we calculate the eigenfunctions of the modes at a Lindblad
resonance in a vertically isothermal disc using the vertical boundary
conditions imposed in the simulations, the flux fractions imparted to
the two-dimensional and r~modes agree with the numbers given above,
indicating that the boundaries are sufficiently distant not to affect
the modes.  At the same time, we discover a further sequence of modes
that fully account for the `missing' $18.9\%$ of the expected flux.
These modes propagate on the same side of the resonance as the \fe
mode but are strongly affected by the boundaries.  They become
spectrally dense in the limit that the boundaries are removed to a
great height, and disappear in the limit $\gamma\to1$.  Most of the
wave energy of these modes is contained away from the mid-plane, near
the boundaries.

We have identified these additional modes as g~modes that are
artificially confined by the boundaries.  In a vertically isothermal
disc with $\gamma>1$, the buoyancy frequency is proportional to $z$
and therefore increases indefinitely with height.  In the absence of
vertical boundaries, internal gravity waves launched at a Lindblad
resonance may propagate with a positive but ever diminishing vertical
group velocity and never reach a turning point.  As a result, they can
never satisfy a standing wave condition and form a vertically confined
g~mode.  However, they constitute a continuous spectrum of outgoing
waves and, as such, receive a non-zero share of the total energy flux.

High in the atmosphere of a real disc, the adiabatic exponent
$\gamma=1$, since the compressional energy of the wave is radiated
away in less than a wave period.  In that case, we expect that some
g~modes will be excited and will be vertically confined.  The
individual g~modes that are excited should take up the required energy
flux in a manner that is independent of the location of any vertical
boundaries.

If we start from our standard polytropic model and continuously reduce
the effective optical thickness towards zero by increasing the
parameter $\beta$ so that the isothermal atmosphere acquires more
mass, the flux fraction for the \fe mode decreases continuously from
$98.3\%$ towards $74.5\%$.

\subsection{The ${\rm p}_1^{\rm e}$ mode}

The ${\rm p}_1^{\rm e}$ mode is launched at a vertical resonance of
the type analysed by Lubow (1981).  In the case of axisymmetric
forcing the resonance condition is $\omega=(\gamma+1)^{1/2}\Omega_z$,
where $\Omega_z$ is the vertical oscillation frequency, equal to
$\Omega$ is a Keplerian disc.  Near the resonance, the ${\rm p}_1^{\rm
  e}$ mode involves a predominantly vertical velocity perturbation
$u_z'\propto z$, corresponding to a `breathing' motion of the disc,
and is excited efficiently by an applied vertical acceleration
$a_z={\cal A}(R)z\exp(-{\rm i}\omega t)$.  We give below an informal
derivation of the wave equation satisfied by the ${\rm p}_1^{\rm e}$
mode near the resonance, in order to predict the energy flux in the
launched wave.  A rigorous derivation can be given, if desired, using
asymptotic expansions.

The exact linearized equations governing axisymmetric Eulerian
perturbations of the disc in the presence of vertical forcing of the
above type are
\begin{equation}
  -{\rm i}\omega u_R'-2\Omega u_\phi'=
  -{{1}\over{\rho}}{{\partial p'}\over{\partial R}}+
  {{\rho'}\over{\rho^2}}{{\partial p}\over{\partial R}},
  \label{u_R'}
\end{equation}
\begin{equation}
  -{\rm i}\omega u_\phi'+
  {{u_R'}\over{R}}{{\partial}\over{\partial R}}(R^2\Omega)+
  u_z'{{\partial}\over{\partial z}}(R\Omega)=0,
\end{equation}
\begin{equation}
  -{\rm i}\omega u_z'=
  -{{1}\over{\rho}}{{\partial p'}\over{\partial z}}+
  {{\rho'}\over{\rho^2}}{{\partial p}\over{\partial z}}+{\cal A}z,
\end{equation}
\begin{equation}
  -{\rm i}\omega\rho'+u_R'{{\partial\rho}\over{\partial R}}+
  u_z'{{\partial\rho}\over{\partial z}}=
  -\rho\left[{{1}\over{R}}{{\partial}\over{\partial R}}(Ru_R')+
  {{\partial u_z'}\over{\partial z}}\right],
  \label{rho'}
\end{equation}
\begin{equation}
  -{\rm i}\omega p'+u_R'{{\partial p}\over{\partial R}}+
  u_z'{{\partial p}\over{\partial z}}=
  -\gamma p\left[{{1}\over{R}}{{\partial}\over{\partial R}}(Ru_R')+
  {{\partial u_z'}\over{\partial z}}\right].
  \label{p'}
\end{equation}
In a thin, Keplerian disc we may take $\Omega=(GM/R^3)^{1/2}$,
neglecting its vertical variation, and set $\partial p/\partial
z=-\rho\Omega^2z$ for vertical hydrostatic equilibrium.  For a
disturbance with radial wavelength short compared to $R$, we may also
neglect the terms involving radial derivatives of $p$ in equation
(\ref{u_R'}), of $\rho$ and of $R$ in equation (\ref{rho'}), and of
$p$ and of $R$ in equation (\ref{p'}).  With these approximations we
obtain, after an integration by parts,
\begin{eqnarray}
  \lefteqn{{\rm i}\omega{\cal A}\int\rho z^2\,{\rm d}z=
  \left[\omega^2-(\gamma+1)\Omega^2\right]
  \int\rho u_z'z\,{\rm d}z}&\nonumber\\
  &&\qquad\qquad\qquad
  -\int\left(\gamma p+z{{\partial p}\over{\partial z}}\right)
  {{\partial u_R'}\over{\partial R}}\,{\rm d}z.
  \label{intrel}
\end{eqnarray}

The quantity in square brackets vanishes at the resonance.  If
$X=R-R_{\rm res}$ is the radial distance from the resonance, we may
apply the linear approximation
\begin{equation}
  \omega^2-(\gamma+1)\Omega^2\approx3(\gamma+1){{\Omega^2}\over{R}}X.
\end{equation}
The leading approximation to the ${\rm p}_1^{\rm e}$ mode near the
resonance is
\begin{eqnarray}
  u_z'&=&{\rm i}\omega zY(X),\nonumber\\
  \rho'&=&\left(\rho+z{{\partial\rho}\over{\partial z}}\right)Y(X),
  \nonumber\\
  p'&=&\left(\gamma p+z{{\partial p}\over{\partial z}}\right)Y(X),
\end{eqnarray}
where $Y(X)$ is a function to be determined.  The horizontal
components of the equation of motion then determine
\begin{equation}
  u_R'=\left({{{\rm i}\omega}\over{\Omega^2-\omega^2}}\right)
  {{1}\over{\rho}}
  \left(\gamma p+z{{\partial p}\over{\partial z}}\right)
  {{{\rm d}Y}\over{{\rm d}X}}.
\end{equation}
The integral relation (\ref{intrel}) then yields the ordinary
differential equation
\begin{equation}
  C_1XY+C_2{{{\rm d}^2Y}\over{{\rm d}X^2}}=C_3,
  \label{de1}
\end{equation}
where the coefficients
\begin{eqnarray}
  C_1&=&3(\gamma+1){{\Omega^2}\over{R}}\int\rho z^2\,{\rm d}z,\nonumber\\
  C_2&=&{{1}\over{\gamma\Omega^2}}\int{{1}\over{\rho}}
  \left(\gamma p+z{{\partial p}\over{\partial z}}\right)^2\,{\rm d}z,
  \nonumber\\
  C_3&=&{\cal A}\int\rho z^2\,{\rm d}z
\end{eqnarray}
are all to be evaluated at the location of the resonance.  By means of
the change of variables
\begin{eqnarray}
  Y(X)&=&C_1^{-2/3}C_2^{-1/3}C_3\,y(x),\nonumber\\
  X&=&C_1^{-1/3}C_2^{1/3}\,x,
\end{eqnarray}
equation (\ref{de1}) is transformed into the inhomogeneous Airy equation,
\begin{equation}
  xy+{{{\rm d}^2y}\over{{\rm d}x^2}}=1,
  \label{airy}
\end{equation}
as also appears in the analysis of Lindblad resonances.  The desired
solution, representing a wave emitted from the resonance and
travelling into $X>0$ with no incoming component, is
\begin{equation}
  y=-\pi\left[{\rm Gi}(-x)+{\rm i}\,{\rm Ai}(-x)\right],
\end{equation}
where ${\rm Ai}$ and ${\rm Gi}$ are homogeneous and inhomogeneous Airy
functions (e.g. Abramowitz \& Stegun 1965).  From the asymptotic form
for large positive $x$,
\begin{equation}
  y\sim-\pi^{1/2}x^{-1/4}\,\exp\left[{\rm i}
  \left({{2}\over{3}}x^{3/2}+{{\pi}\over{4}}\right)\right],
\end{equation}
we determine the radial energy flux (averaged in time, and integrated
azimuthally and vertically) in the wave at some distance from the
resonance,
\begin{eqnarray}
  {\cal F}&=&\pi R\int{u_R'}^*p'\,{\rm d}z,\nonumber\\
  &=&{{\pi^2R^2{\cal A}^2}\over{3\omega}}\int\rho z^2\,{\rm d}z.
  \label{linear_flux_p}
\end{eqnarray}
This is directly analogous to equation (\ref{gt}) for a Lindblad
resonance.

\subsection{The \foro mode}

Generically, the ${\rm f}^{\rm o}$ mode, which effects a corrugation
or tilt of the disc, is launched at a vertical resonance.  In the case
of axisymmetric forcing the resonance condition is $\omega=\Omega_z$.
In a Keplerian disc this coincides with the Lindblad resonance and the
\foro mode is launched at a hybrid vertical/Lindblad resonance that
has not been discussed in the literature.

Near the resonance, the \foro mode involves a vertical velocity
perturbation $u_z'$ independent of $z$ {\it and\/} horizontal velocity
perturbations $u_R',u_\phi'\propto z$ of comparable magnitude.  It is
excited efficiently by an applied vertical acceleration $a_z={\cal
  A}(R)\exp(-{\rm i}\omega t)$.  We give below an informal
derivation of the wave equation satisfied by the \foro mode near the
resonance.  Again, a rigorous derivation can be given using asymptotic
expansions.

The equations governing the perturbations are the same as in the
previous subsection except that the vertical acceleration no longer
contains the factor $z$.  We now obtain the integral relation
\begin{equation}
  {\rm i}\omega{\cal A}\int\rho\,{\rm d}z=
  (\omega^2-\Omega^2)\int\rho u_z'\,{\rm d}z-
  \int{{\partial p}\over{\partial z}}
  {{\partial u_R'}\over{\partial R}}\,{\rm d}z,
  \label{intrel2}
\end{equation}
in which the quantity in brackets vanishes at the resonance.

If $X=R-R_{\rm res}$ is the radial distance from the resonance, we
may apply the linear approximation
\begin{equation}
  \omega^2-\Omega^2\approx{{3\Omega^2}\over{R}}X.
  \label{linear_approximation}
\end{equation}
The leading approximation to the \foro mode near the
resonance is
\begin{eqnarray}
  u_z'&=&{\rm i}\omega Y(X),\nonumber\\
  \rho'&=&{{\partial\rho}\over{\partial z}}Y(X),\nonumber\\
  p'&=&{{\partial p}\over{\partial z}}Y(X).
\end{eqnarray}
The horizontal components of the equation of motion then determine
\begin{equation}
  u_R'=\left({{{\rm
i}\omega}\over{\Omega^2-\omega^2}}\right){{1}\over{\rho}}
  {{\partial p}\over{\partial z}}{{{\rm d}Y}\over{{\rm d}X}}.
\end{equation}
The denominator of this expression may be represented by the same
linear approximation, equation (\ref{linear_approximation}).  The
integral relation (\ref{intrel2}) then yields the ordinary
differential equation
\begin{equation}
  C_1XY+C_2{{{\rm d}}\over{{\rm d}X}}
  \left({{1}\over{X}}{{{\rm d}Y}\over{{\rm d}X}}\right)=C_3,
  \label{de2}
\end{equation}
where the coefficients
\begin{eqnarray}
  C_1&=&{{3\Omega^2}\over{R}}\int\rho\,{\rm d}z,\nonumber\\
  C_2&=&{{R}\over{3\Omega^2}}\int{{1}\over{\rho}}
  \left({{\partial p}\over{\partial z}}\right)^2\,{\rm d}z,
  \nonumber\\
  C_3&=&{\cal A}\int\rho\,{\rm d}z
\end{eqnarray}
are all to be evaluated at the location of the resonance.  By means of
the change of variables
\begin{eqnarray}
  Y(X)&=&C_1^{-3/4}C_2^{-1/4}C_3\,y(x),\nonumber\\
  X&=&C_1^{-1/4}C_2^{1/4}\,x,
\end{eqnarray}
equation (\ref{de2}) is transformed into
\begin{equation}
  xy+{{{\rm d}}\over{{\rm d}x}}
  \left({{1}\over{x}}{{{\rm d}y}\over{{\rm d}x}}\right)=1.
  \label{fresnel}
\end{equation}
Unlike equation (\ref{airy}), this equation allows wave propagation on
both sides of the resonance, $x>0$ and $x<0$.  The complementary
functions of equation (\ref{fresnel}) are
$\cos({\textstyle{{1}\over{2}}}x^2)$ and
$\sin({\textstyle{{1}\over{2}}}x^2)$, and the general solution of the
inhomogeneous equation can therefore be expressed in terms of Fresnel
integrals.  The desired solution, representing one wave travelling
into $X>0$ and another wave travelling into $X<0$, with no incoming
component, is
\begin{eqnarray}
  y&=&\left[-{{1}\over{2}}{\rm i}\pi^{1/2}-
  \int_0^x\sin({\textstyle{{1}\over{2}}}t^2)\,{\rm d}t\right]
  \cos({\textstyle{{1}\over{2}}}x^2)\nonumber\\
  &&+\left[-{{1}\over{2}}{\rm i}\pi^{1/2}+
  \int_0^x\cos({\textstyle{{1}\over{2}}}t^2)\,{\rm d}t\right]
  \sin({\textstyle{{1}\over{2}}}x^2).
\end{eqnarray}
From the asymptotic form for large positive $x$,
\begin{equation}
  y\sim-\left({{\pi}\over{2}}\right)^{1/2}\,\exp\left[{\rm i}
  \left({{1}\over{2}}x^2+{{\pi}\over{4}}\right)\right],
\end{equation}
we determine the energy flux in the wave at some distance from the
resonance on the positive side,
\begin{eqnarray}
  {\cal F}&=&\pi R\int{u_R'}^*p'\,{\rm d}z,\nonumber\\
  &=&{{\pi^2R^2{\cal A}^2\Sigma}\over{6\omega}}.
  \label{linear_flux_foro}
\end{eqnarray}
This is exactly half the value given by equation (\ref{gt}) for a
Lindblad resonance.  From the symmetry property $y(-x)=-y^*(x)$, it
follows that the wave emitted into $x<0$ carries an equal and
oppositely directed energy flux.  Therefore the total energy flux is
identical to that for a Lindblad resonance, although of course ${\cal
  A}$ is the amplitude of the vertical, not horizontal, acceleration.

\end{document}